\documentclass[12pt]{article}
\usepackage{amsmath,amsthm,amsfonts}
\usepackage{graphicx,psfrag,epsf}
\usepackage{enumerate}
\usepackage{natbib}
\usepackage{url}
\usepackage{bm}
\RequirePackage[colorlinks,citecolor=blue,urlcolor=blue]{hyperref}
\usepackage{algorithm}
\usepackage{algpseudocode}
\usepackage{algorithmicx}
\usepackage{multirow}
\usepackage{adjustbox}
\usepackage{tabularx}
\usepackage{booktabs}
\usepackage{subcaption}
\usepackage[dvipsnames]{xcolor}
\usepackage{xfrac}
\usepackage{mathrsfs}

\def\bx{\mathbf{x}}

\def\bgam{{\bm \gamma}}
\def\balp{{\bm \alpha}}
\def\bth{{\bm \theta}}
\def\blam{\boldsymbol{\lambda}}

\def\uI{\mathrm{I}}
\def\Be{\text{Be}}

\def\E{\text{E}}
\def\var{\text{var}}

\def\waic{\text{WAIC}}
\def\tplus{t\!+}
\def\tminus{t\!-}

\theoremstyle{remark}

\newcommand{\note}[1]{#1} 

\theoremstyle{plain}
\newtheorem{proposition}{Proposition}
\newcommand{\hypgeo}[2]{%
  {\vphantom{F}}_{#1}\kern-\scriptspace F_{#2}%
}

\usepackage{setspace}
\def\spacingset#1{\renewcommand{\baselinestretch}%
{#1}\small\normalsize} 
\spacingset{1}

\begin{document}

\title{\bf Density Discontinuity Regression:\\
   Modeling Heterogeneous Sorting at Thresholds}
  \author{Surya T Tokdar\thanks{Address for Correspondence: 214 Old Chemistry Bldg, Duke University, Durham NC 27708. E-mail: surya.tokdar@duke.edu. }\hspace{.2cm}\\
    Department of Statistical Science, Duke University\\
    Rik Sen\\
    Department of Finance, University of Georgia\\
    Haoliang Zheng\\
    Department of Statistics \& Data Science, National University of Singapore\\
    and\\
    Shuangjie Zhang\\
    Department of Statistics,  Texas A\&M University}
    \date{}
    
\maketitle
  
\bigskip
\begin{abstract}
A jump discontinuity in the density of an economic variable at a known threshold may signal strategic manipulation by relevant actors. This paper introduces density discontinuity regression for estimating covariate effect on the jump size, adjusting for effects on near-threshold density shape and accommodate sign constraints for one-sided manipulation. Improved bias--variance tradeoff is gained by adaptively localizing estimation near the threshold. Application to U.S. corporate voting reveals strategic manipulation around the majority threshold is stronger under Institutional Shareholder Services opposition, weaker with high analyst monitoring, and conditionally stronger for larger firms; these insights are missed by exploratory analyses.
\end{abstract}

\noindent%

\spacingset{1.25} 

\section{Introduction}
\label{sec:intro}

\subsection{Why density discontinuity regression?}
Many policy designs trigger a binary outcome when a continuous policy variable crosses a known threshold. Individuals or firms aware of the thresholding rule often adjust their behavior to stay on the favorable side of the cutoff. Such strategic adjustments are hard to measure at the individual level but could manifest at the population level in the form of a density discontinuity: many more individuals landing on one side of the threshold relative to the other. 
For instance, in shareholder voting on management--sponsored proposals, close votes are disproportionately resolved just above the majority threshold
\citep{listokin2008management, babenko2026management}, while shareholder--sponsored proposals display the mirror-image pattern, with an excess of narrow failures \citep{bach2019close}.
In household income data, the distribution of the wife's share of income exhibits a sharp drop just to the right of one-half, where the wife would earn more than the husband \citep{bertrand2015gender}. In hedge fund data, small reported gains are much more common than small reported losses \citep{bollen2009hedge}. These examples differ in institutional details, but they share the same empirical signature: the density of the running variable changes discontinuously at a threshold which agents have incentives to avoid, attain, or cross \citep[see also][]{kleven2016bunching, jales2017identification}.

Existing density-discontinuity analyses focus mainly on the detection of a jump at the threshold \citep{mccrary2008manipulation, otsu2013estimation, cattaneo2020simple}. In many economic applications, however, the important question comes next: what measurable factors are associated with the existence and the size of the jump? Identifying these factors is the first step toward understanding what drives the underlying strategic behavior. For instance, in a moderately large sample of management-sponsored proposals from U.S.\ corporations, the density of the shareholder vote share can be seen to exhibit a clear and statistically significant jump at the majority threshold (Figure~\ref{fig:f1-marginal}). This observation aligns with the understanding that managements typically have both the motive and the means to engineer narrow wins \citep{listokin2008management, babenko2026management}. But the aggregate test obscures what governance researchers want to know. Is the jump larger when Institutional Shareholder Services (ISS) recommends against management, signaling a credible threat to the proposal \citep{malenko2016role}? Is it smaller when analyst coverage disciplines management behavior, as the monitoring literature would predict \citep{chen2015analysts}? Does firm size attenuate or amplify the apparent manipulation, given competing arguments about scale, capacity, and visibility \citep{aggarwal2019power}? 
Segmenting the sample one covariate at a time---the standard exploratory approach, and the one taken in the applied literature on this setting---points to heterogeneity along some of these dimensions but is silent or misleading on others. It cannot deliver the partial effect of each covariate holding the rest fixed, which is what these questions require, and it is least reliable precisely when the covariates are correlated with one another, as they are here.

This methodological gap is precisely what we attempt to fill in this paper with a novel {\it density discontinuity regression} analysis tool which estimates covariate effects on the jump size at the threshold while adjusting for the local density shape.
%
%
We propose a regression framework that estimates how the log-density ratio at the threshold $t$,
\begin{equation}
j(\bx) :=\; \log \frac{f(\tplus | \bx)}{f(\tminus | \bx)} \;=\; g(\bx'\balp),
\label{eq:disc}
\end{equation}
varies with covariates $\bx$ as dictated by unknown coefficients $\balp$. The target $j(\bx)$ is the same quantity that McCrary-style tests examine in the aggregate, now expressed as a regression on $\bx$ so that standard applied-econometrics questions---partial effects, sign tests, model comparison---become readily addressable. Our estimation framework embeds sign constraint on the jump, such as $j(\bx) \ge 0$ while allowing it to be exactly zero (no jump) for some but not all values of $\bx$, and jointly estimates $\balp$ along with the the shape of the conditional density $f(y|\bx)$ near the threshold. These choices are motivated as follows. 

\subsection{Existing approaches}
To the best of our knowledge, the only work to offer a regression analysis of $j(\bx)$ is \citet{goncharov2023central}, who trim the sample to $y \in [t-\Delta, t+\Delta]$ and regress the binary indicator $y^* = \uI(y \ge t)$ on covariates by ordinary least squares. A natural refinement is logistic regression, since 
\begin{equation}
\log\frac{\Pr(y^*=1| \bx)}{\Pr(y^*=0| \bx)} 
 = \log \frac{\int_t^{t + \Delta} f(y|\bx)dy}{\int_{t-\Delta}^{t} f(y|\bx)dy} \approx j(\bx)~\text{for small $\Delta$}.
\label{eq:bor}
\end{equation} 
This binary outcome regression (BOR) analysis has two weaknesses that matter in practice. First, the approximation error in \eqref{eq:bor} is governed by the slope of $f(y|\bx)$ near $t$. The larger the slope, the more it biases the estimate of $\balp$. Bias could be reduced by decreasing $\Delta$, but the optimal mean squared error rate cannot be better than $O(n^{-2/3})$ (Section~\ref{sec:theory}). Choosing $\Delta$ to navigate this bias--variance tradeoff is essential, but BOR offers no principled guide because it does not offer any model for the data generating procedure for the outcome variable $y$. Second, BOR places no restriction on the sign of $j(\bx)$, and its estimates of $\balp$ can be hard to interpret as a result. In our application to corporate proposal voting, BOR attaches negative point estimates to both ``ISS against recommendation'' and firm size (Table~\ref{tab:vote-simple}). The first runs against the direction that governance reasoning suggests and that the broader subsample evidence supports \citep{babenko2026management}. More tellingly, the BOR intercept in that application is inflated by an amount very close to the bias our rate analysis predicts for it, as we show in Section \ref{sec:voting}.

A complementary line of work estimates $f(t{\pm}| \bx)$ nonparametrically---for instance, via the boundary-adaptive local-polynomial conditional density estimator of \citet{cattaneo2024boundary} applied separately on each side of the threshold---and takes the log-ratio pointwise. This approach is the right tool when interest lies in pointwise estimation of $j(\bx)$ across $\bx$, and it has well-developed large-sample theory, MSE-optimal bandwidths, and bias correction. 
Our inferential target is different: we want a coefficient vector $\balp$ that isolates the partial effect of each covariate on the jump size, adjusted for the others, with standard regression-style summaries such as sign tests and confidence intervals. Pointwise estimation of $j(\bx)$ does not directly deliver this, and projecting such estimates onto covariates in a second step would compound estimation error without the efficiency gains that come from joint modeling.

\cite{byzalov2019modeling}  pursue a related goal in the earnings management literature, modeling how covariates affect the probability that a firm manipulates earnings across a threshold. Their estimand---the conditional manipulation probability---measures only the prevalence of meet-or-just-beat behavior as a function of firm characteristics. Our estimand targets both prevalence (is $j(\bx) \ne 0$?) and magnitude (size of $j(\bx)$) of potential manipulative behavior, expressed as a regression on covariates, and helps isolate how each covariate shifts the relative density on either side of the cutoff without requiring assumptions about the mechanics of the manipulation process itself. A further difference concerns window selection. As in \citet{goncharov2023central}, \cite{byzalov2019modeling} acknowledge the absence of any formal criterion for choosing the estimation interval and rely instead on visual inspection and sensitivity checks over a modest range of alternatives. Our predictive fit--based adaptive localization procedure selects the trimming window from the data, providing a principled and replicable answer to a choice that would otherwise be left to researcher judgment.

\subsection{A new approach} 
Any density function $f(y|\bx)$ which has a jump discontinuity of size $j(\bx)$ at $y = t$ but is continuous otherwise could be factorized as
\begin{equation}
f(y| \bx) = \frac{1}{Z(\bx)} \, b(y| \bx) \cdot e^{-\uI(y<t)\cdot j(\bx)},
\label{eq:decomp}
\end{equation}
where $b(y|\bx)$ is a continuous density function and $Z(\bx) = \int b(y| \bx) e^{-\uI(y<t)\cdot j(\bx)} dy$ is the appropriate normalizing constant (simply take $b(y|\bx) \propto f(y|\bx) [\frac{f(t+|\bx)}{f(t-|\bx)}]^{\uI(y < t)}$). Therefore, statistical modeling and estimation of $f(y|\bx)$ could separately target the two pieces, the jump size $j(\bx)$ and the continuous density $b(y|\bx)$ which we label the {\it baseline} density because, in the absence of any strategic adjustments (i.e., $j(\bx) = 0$), we would have the exact identity $f(y|\bx) = b(y|\bx)$. We propose a statistical analysis under the parametric choices: 
\begin{align*}
j(\bx)& = (\bx'\balp)_+ ~\text{is as in \eqref{eq:disc} with}~ g(z) = z_+,~\text{and}\\
b(y|\bx)& ~\text{is the density function of}~\Be(\psi_1(\bx;\bgam), \psi_2(\bx;\bgam)). 
\end{align*}
The choice of  $g(z) = z_+$ forces $j(\bx) \ge 0$ and is appropriate in our voting application and in most bunching settings where manipulation is one sided (tax brackets, regulatory cutoffs, grade thresholds, etc.). However, assuming $b$ is in the beta parametric family is clearly restrictive. Although a nonparametric estimation of $b(y|\bx)$ could give added flexibility, a joint estimation of $\balp$ and $b$ appears complicated in the classical nonparametric sense, and computationally challenging for Bayesian nonparametric estimation due to the presence of the covariate-dependent normalizing constants $Z(\bx)$ (more in Section \ref{sec:discussion}). With the parametric beta model for $b$, $Z(\bx)$ remains easy to compute so that a tractable likelihood-based inference can be carried out.

Crucially, to mitigate any bias introduced by the parametric restriction on $b$, we localize our estimation to samples with $y \in [t -\Delta, t + \Delta]$, where the trimming window $\Delta$ is chosen by a WAIC-based model selection criterion \citep{watanabe2013widely}. Although the localization theme is similar to BOR, the parametric modeling of $b$ has two clear advantages: the mean squared error rate in estimating $\balp$ improves to $n^{-(2r+2)/(2r+3)}$, where $r \ge 0$ indexes the agreement between the slope and higher order derivatives of $b(y| \bx)$ at $t$ and what the parametric family allows (matching BOR rate of $n^{-2/3}$ at $r=0$ and approaching the parametric rate $n^{-1}$ as $r$ grows, though the two-parameter beta family caps it at $n^{-6/7}$ with $r = 2$; Section~\ref{sec:theory}), and the WAIC-based choice of $\Delta$ is a fully automated, data-dependent procedure with good adaptive properties in moderate to large samples (rather than by inspection as in BOR). 
Applied to the proposal-voting data, the framework delivers partial effects that marginal segmentation cannot isolate and that BOR estimates with the wrong sign.
The jump at the majority threshold is sharply larger when ISS recommends against management ($\hat\alpha_{\text{ISS}} = 0.28$, 95\% CI $[0.16, 0.40]$), smaller when analyst coverage is high ($-0.48$, $[-0.68, -0.30]$), and---once analyst coverage is held constant---larger for bigger firms ($0.22$, $[0.04, 0.40]$). The firm-size effect is ambiguous in the marginal analysis, and BOR estimates it to be negative but insignificant. These differences point to the the possibility that both adjustment by other covariates and a restriction of the sign of $j(\bx)$ have substantial impact on the analysis. Past stock returns and Tobin's $Q$ show no detectable association after adjustment. Estimated jumps imply that manipulation is most pronounced for ISS-opposed proposals from large, lightly covered firms, where the estimated density ratio at the threshold can exceed 20.

\begin{table}[!t]
    \caption{Estimates of $\balp_0$ from BOR analyses of corporate voting ($\Delta = \frac1{10}, n_\Delta = 1598$). 
    \label{tab:vote-simple}}
    \centering
\begin{tabular}{c|rrrrrr}
\toprule
Method   & \multicolumn{6}{c}{Estimate of regression coefficients}\\[6pt]
       & Intercept & ISS & Analyst & Return & Q-value & Size\\
\hline
Logistic  &  1.66    & --0.08  &   --0.29 & 0.15 & 0.05 & --0.05\\[-6pt]
& {\tiny (1.50,1.82)} & {\tiny (--0.18, 0.02)} & {\tiny (--0.47, --0.11)} & {\tiny (0.00,  0.30)} & {\tiny (--0.11,  0.22)} & {\tiny (--0.23,  0.12)}\\
Least-squares & 0.83     & --0.01  &   --0.04 &  0.02 & 0.01 & -0.01\\[-6pt]
& {\tiny (0.81,0.85)} & {\tiny (--0.02, 0.00)} & {\tiny (--0.06, --0.01)} & {\tiny (0.00,  0.04)} & {\tiny (--0.02,  0.03)} & {\tiny (--0.03,  0.02)}\\
\bottomrule
\end{tabular}
\end{table}

\subsection{Contributions} 
We summarize three main contributions this article makes. First, we define a regression estimand for density discontinuities: covariate effects on the log-density ratio at a known threshold. This estimand is the natural conditional analogue of the McCrary-style density jump. Second, we develop an estimator that jointly models the jump and the local density shape, which avoids the main weakness of binary (above-threshold) outcome regressions: they mix variation in manipulation with variation in the smooth component of the density. Third, we introduce a data-adaptive localization procedure and allow the researcher to impose sign restrictions when the institutional setting implies one-sided manipulation. Applied to shareholder voting, the method shows that apparent strategic sorting is stronger when ISS opposes management, weaker when analyst coverage is high, and stronger for larger firms. The remainder of the paper is organized as follows. Section~\ref{sec:theory} develops the bias--variance rate analysis. Section~\ref{sec:betareg} specifies the beta regression model and Bayesian estimation under the non-negativity constraint and introduces adaptive choice of the trimming window via WAIC. The statistical performance of fixed and adaptive localization based parametric estimation are studied in detail in Section~\ref{sec:betareg-num}. Section~\ref{sec:voting} reports the corporate voting analysis. Section~\ref{sec:discussion} discusses extensions and limitations.


\section{Why local parametric modeling helps}
\label{sec:theory}

This section gives the theoretical intuition behind our recommendation to model the local density rather than work with a binary outcome indicator. Assume that the true (continuous) baseline density is $b_0(y|\bx)$ and the jump size relates to the covariates as $j(\bx) = \bx'\balp_0$. Our analysis is carried out under the additional assumption that $b_0(y|\bx)$ is differentiable at $y = t$. The fundamental idea is that by estimating and utilizing the slope and higher order derivatives of $b_0(y|\bx)$ at $y = t$, one can enforce a more favorable bias-variance tradeoff in the estimation of $\balp_0$ than BOR. 

BOR trims the sample to $y \in [t-\Delta, t+\Delta]$ and regresses $y^* = \uI(y \ge t)$ on $\bx$. From \eqref{eq:bor}, by Taylor's approximation,
\begin{align*}
\log \frac{\Pr(y^*=1|\bx)}{\Pr(y^*=0|\bx)} & = \log \frac{f_0(\tplus |\bx)}{f_0(\tminus |\bx)} + \frac{\Delta}{2} \times \left\{\frac{\dot f_0(\tplus | \bx)}{f_0(\tplus | \bx)} + \frac{\dot f_0(\tminus | \bx)}{f_0(\tminus | \bx)} \right\} + o(\Delta)\\
 & = \bx'\balp_0 + \Delta \cdot m_{0,1}(\bx) + o(\Delta)
\end{align*}
where $m_{0,l}(\bx) := \frac{\partial^l}{\partial y^l} \log b_0(y|\bx)|_{y = t}, l \ge 1$. Suppose $b_0$ has the slope $m_{0,1}(\bx) = \bx'\blam_0 + o(\|\bx\|)$ for some $\blam_0$. Then the right hand side of the last display equals $\bx'(\balp_0 + \Delta\blam_0) + o(\Delta)$, and thus the BOR
estimate of $\alpha_j$ would have a bias $\approx \Delta \lambda_{0,j}$, which grows linearly in $\Delta$. The variance of the estimate is $O(1/(n\Delta))$ since the trimmed sample has effective size $n\Delta$. The mean squared error is therefore 
\[
(\text{BOR})\qquad
E[(\hat\alpha_j - \alpha_{0,j})^2] \asymp \Delta^2 + (n\Delta)^{-1}
\] up to constants, minimized at $\Delta \propto n^{-1/3}$ and giving rate $n^{-2/3}$. This rate is the best BOR can achieve unless the true baseline density is exactly flat at $y = t$ (i.e., $\blam_0 = 0$).

Next consider the case where we estimate $b_0(y|\bx)$ by modeling it as a member of some parametric family of smooth densities $\mathscr{B} = \{b(y|\bx;\bgam), \bgam \in \Gamma \subset \mathbb{R}^k\}$, and localize the estimation to samples with $y \in [t - \Delta, t + \Delta]$, with the (localized) maximum likelihood estimator is given by
\[
(\hat\bgam,\hat\balp) = \arg\max_{\bgam,\balp} \prod_{i: |y_i-t|\le \Delta} f^\Delta(y_i|\bx_i;\bgam,\balp),
\]
where $f^\Delta(y|\bx;\bgam,\balp)$ is the restriction of $f(y|\bx;\bgam,\balp) \propto b(y|\bx;\bgam) \exp\{-(\bx'\alpha) \cdot I(y < t)\}$ to the interval $[t - \Delta, t + \Delta]$. Let $m_l(\bx;\bgam) = \frac{\partial^l}{\partial y^l} \log b(y|\bx;\bgam)|_{y = t}, l \ge 1$, denote the derivatives of the model baseline density at the threshold, and suppose $r \ge 0$ is the largest integer such that there is a $\bgam_0 \in \Gamma$ with $m_l(\bx; \bgam_0) = m_{0,l}(\bx)$, $l = 1,\ldots,r$. Then, under appropriate regularity conditions, the bias and variance of $\hat\alpha_j$ should equal, respectively, $O(\Delta^{r+1})$ and $O((n\Delta)^{-1})$, and thus one should have
\[
(\text{Parametric})\qquad E[(\hat\alpha_j - \alpha_{0,j})^2] \asymp \Delta^{2(r+1)} + (n\Delta)^{-1}.
\]
which admits an optimal bias-variance tradeoff at $\Delta \propto n^{-\frac{1}{2r+3}}$ giving the best possible error rate of $n^{-\frac{2r+2}{2r+3}}$; see Appendix~\ref{app:rate} for a formal statement and an informal proof sketch.

The magnitude of this accuracy gain is governed by a single index $r \ge 0$ measuring how many derivatives of $\log b_0(y \mid \bx)$ at $y = t$---viewed as functions of $\bx$---lie in the family generated by $\bgam$. The case $r = 0$ corresponds to no match, and recovers the BOR rate. Larger $r$ corresponds to richer agreement between the family and the truth at $y=t$, with the parametric rate $n^{-1}$ as the conceptual ceiling at $r = \infty$. How large an $r$ is achievable in practice is a property of the parametric family chosen for baseline density and is, in general, finite. For the beta family used in Section~\ref{sec:betareg}, we show below that the achievable ceiling is $r = 2$, giving a best-case rate of $n^{-6/7}$. 

We reiterate here that the gain over BOR is not free. Mitigating bias by estimating the shape of $b(y|\bx)$ near $y=t$ comes at an increased computational cost. A practicable but completely nonparametric estimation of the baseline density remains beyond the reach of current statistical literature (Section \ref{sec:discussion}). The strategy we adopt in Section~\ref{sec:betareg} is to use a flexible but tractable parametric family---beta regression with covariate-dependent shape parameters---and to localize the analysis to a window chosen by the data (Section~\ref{sec:adapt}). Localization mitigates misspecification: even if the family fails to capture $b_0$ globally, it need only match the first few derivatives at $t$ to deliver a faster rate than BOR.

Finally, it should not be overlooked that the rate calculation is only a heuristic guide rather than a fully rigorous theorem---we have not verified that the localized maximum likelihood estimator's asymptotic variance is uniform in $\Delta$ as $\Delta \to 0$ and $n \to \infty$, and we have not characterized the constants. Additionally, the methodology we develop in Section \ref{sec:betareg} is not based on maximum likelihood estimation. We use Bayesian estimation of $(\balp, \bgam)$ necessitated partly by the fact that the sign-constrained jump formulation makes the likelihood function irregular and partly by the need to determine a data-adaptive localization window $\Delta$ by framing the problem as a model selection question. However, the numerical studies of Sections~\ref{sec:betareg-num} bear out the qualitative implications: localization improves estimation when the local model is misspecified, the optimal window is data-dependent, and the resulting estimator dominates BOR on both bias and coverage.

\section{Beta regression with discontinuity}
\label{sec:betareg}
\subsection{Exact model and complete likelihood}
\label{sec:model}
An important advantage of choosing $b(y|\bx)$ to be the $\Be(\psi_1(\bx;\bgam), \psi_2(\bx;\bgam))$ density is that one could write the normalizing constant in \eqref{eq:decomp} as 
\[
Z(\bx) = c(t; j(\bx), \psi_1(\bx;\bgam), \psi_2(\bx;\bgam)), 
\]
where, for $t \in (0,1), j \ge 0, \psi_1 > 0, \psi_2 > 0$, 
\[
c(t;j,\psi_1,\psi_2) = \int_0^1 y^{\psi_1-1}(1-y)^{\psi_2-1}e^{-j\cdot I(y < t) } dy = B(\psi_1,\psi_2) \times \{1 - (1 - e^{-j})I_t(\psi_1, \psi_2)\}
\] 
with $B(\psi_1, \psi_2) = \int_0^1 y^{\psi_1-1}(1-y)^{\psi_2-1}dy$ denoting the beta function and $I_t(\psi_1, \psi_2) = \int_0^t y^{\psi_1-1}(1-y)^{\psi_2-1}dy/B(\psi_1, \psi_2)$ denoting the regularized incomplete beta function. It is important to appreciate that in spite of the introduction of a jump discontinuity, the normalizing constant  is available in essentially closed form; see Appendix \ref{app:a} for discussion on its fast evaluation. 

For the beta shape functions, we consider $\psi_i(\bx; \bgam) = s(\bx'\bgam_i)$ where $s(z) = a + (b-a)\cdot e^z/(1 + e^{z})$ is the monotonically increasing sigmoid function bounded between $a$ and $b$, with $0  < a < b < \infty$. For our applications and numerical studies we take $a = 0.1$ and $b = 30$. These wide bounds allow essentially the full range of beta density shapes but help avoid numerical overflow issues that may occur while computing with beta distributions with shape parameters that are either very close to zero or very large. Consequently, our parametric model for the outcome density equals
\begin{equation}
f(y|\bx) = \frac{y^{s(\bx'\bgam_1)-1}(1-y)^{s(\bx'\bgam_2)-1} \times e^{-I(y < t)\cdot (\bx'\balp)_+}}{c(t; (\bx'\balp)_+, s(\bx'\bgam_1), s(\bx'\bgam_2))},~~y \in (0,1),
\label{eq:par}
\end{equation}
with model parameter $\bth = (\bgam_1,\bgam_2,\balp) \in \mathbb{R}^{3p}$. Without any localization, the complete likelihood function in $\bth$ can be written as
\begin{equation}
L(\bth) \propto \prod_{i = 1}^n \frac{y_i^{s(\bx_i'\bgam_1)}(1-y_i)^{s(\bx_i'\bgam_2)} \cdot e^{-I(y_i < t)\cdot (\bx_i'\balp)_+}}{c(t; (\bx_i'\balp)_+, s(\bx_i'\bgam_1), s(\bx_i'\bgam_2))}
\label{eq:lik}
\end{equation}
where, $\{(\bx_i, y_i), 1 \le i \le n\}$ is the entire observed sample.

\subsection{Localization}
\label{sec:localization}

 As discussed in Section \ref{sec:theory}, localizing the data near the threshold of interest allows one to mitigate bias, since estimation accuracy depends only on the agreement between derivatives $m_l(\bx)$ of $\log b(y|\bx)$ at $y = t$ and those of the true baseline function. For the beta family,
\begin{equation}
m_l(\bx;\bgam) = (l-1)!\left[(-1)^{l-1}t^{-l}\{s(\bx'\bgam_1) - 1\} - (1-t)^{-l}\{s(\bx'\bgam_2) - 1\}\right],
\label{eq:beta-derivs}
\end{equation}
and in particular, with $t = 0.5$, $m_1(\bx;\bgam) = 2\{s(\bx'\bgam_1) - s(\bx'\bgam_2)\}$, $m_2(\bx;\bgam) = 8 - 4\{s(\bx'\bgam_1) + s(\bx'\bgam_2)\}$, $m_3(\bx;\bgam) = 16\{s(\bx'\bgam_1) - s(\bx'\bgam_2)\} = 8\, m_1(\bx;\bgam)$, and so on. Each $m_l$ is a fixed linear combination of $s(\bx'\bgam_1)$ and $s(\bx'\bgam_2)$---two scalar functions of $\bx$---so the family carries exactly two independent local degrees of freedom at $y = t$, regardless of how flexibly $\bgam_1$ and $\bgam_2$ are parameterized. The consequence for the rate analysis of Section~\ref{sec:theory} is that, generically, the beta family can match the truth's first two derivatives at $y = t$ but not a third independent one. The corresponding ceiling under Proposition~\ref{prop:rate} is the rate $n^{-6/7}$, strictly better than the BOR rate $n^{-2/3}$ but strictly worse than the parametric rate $n^{-1}$. The same conclusion holds for general $t \in (0,1)$; the two-degree-of-freedom ceiling is a property of the family rather than of the threshold value.

When fitting the parametric model \eqref{eq:par} only to samples with $y_i \in [t - \Delta, t + \Delta]$, the likelihood function \eqref{eq:lik} must be modified in two ways:
\begin{align}
L(\bth; \Delta) & =  \prod_{i: |y_i - t| \le \Delta} \frac{f(y_i | \bx_i, \bth)}{\int_{t - \Delta}^{t + \Delta} f(y | \bx_i, \bth) dy} \nonumber\\
& \propto  \prod_{i: |y_i - t| \le \Delta} \frac{y_i^{s(\bx_i'\bgam_1)}(1-y_i)^{s(\bx_i'\bgam_2)} \cdot e^{-I(y_i < t)\cdot (\bx_i'\balp)_+}}{c(t; (\bx_i'\balp)_+, s(\bx_i'\bgam_1), s(\bx_i'\bgam_2), [t \pm \Delta])},
\label{eq:trunc-lik}
\end{align}
where the product is now restricted to the localized samples and the density $f(y|\bx)$ must be renormalized to the local interval. Again, the advantage of choosing a beta family as the baseline becomes clear, as the normalizing constants could be calculated by evaluating the expressions
\[
c(t;j,\psi_1, \psi_2, [t_1,t_2]) = B(\psi_1, \psi_2) \times \{I_{t_2}(\psi_1, \psi_2) - e^{-j}I_{t_1}(\psi_1, \psi_2) - (1- e^{-j})I_t(\psi_1, \psi_2)\},
\]
for any $t_1 < t < t_2$. Although these truncation based normalizing constants are computationally more expensive, in Appendix \ref{app:a} we discuss computing strategies to mitigate this problem.

\subsection{Bayesian parameter estimation}

For our sign-constrained jump formulation $j(\bx) = (\bx'\balp)_+$, the likelihood function is flat in $\balp$ over an entire region of $\balp$ values with $\max_i \bx_i'\balp \le 0$. Such irregularly shaped likelihood functions pose considerable computational challenges to calculating the maximum likelihood estimate, and more pronouncedly, to estimating its standard error. Such problems could be alleviated with mild regularization by introducing a prior distribution toward Bayesian estimation aided by Markov chain Monte Carlo (MCMC) computing. 
%
Bayesian estimation also proves useful in numerically comparing model fits at various levels of truncation. Effective model comparison requires a careful balancing between how well a candidate model fits the data against how complex the model is. For models that vary only in truncation level, a closed form expression of model complexity, such as those used by Akaike information criterion, is not readily available. But the WAIC model scoring method adopted here provides an easy numerical alternative which works seamlessly with models fitted with the MCMC algorithm. We will come back to this topic in Section \ref{sec:adapt}.

Assume that the $n\times p$ design matrix of predictors, $X = [\bx_1,\ldots,\bx_n]'$ has 1 on its first column and its remaining columns are scaled to have zero sample mean and unit sample variance each. As a default prior specification, we take the coordinates of $\balp$ to be independent centered normal variables with variance 1 and the coordinates of each $\bgam_i$ vector to be independent centered normal variables with variance $1/p$ each; with independence assumed between $\balp$, $\bgam_1$, and $\bgam_2$. Consequently, the joint posterior density of the model parameters could be written as $p(\bth | \mbox{data}) \propto L(\bth)\times e^{-\frac{\|\balp\|^2 + p(\|\bgam_1\|^2 + \|\bgam_2\|^2)}{2}}$. This posterior distribution could be numerically summarized by means of MCMC, where one runs a Markov chain sampler, with $p(\bth|\mbox{data})$ as the stationary distribution, to make sequential draws of $\bth$. First several draws are discarded and the remaining draws are thinned to produce a final representative sample of values $\{\bth_1,\ldots,\bth_M\}$. While a number of different samplers are available to carry out the sampling, we have found it efficient to use an elliptical slice sampler \citep{murray2010elliptical, nishihara2014parallel} on the $3p$-dimensional space of $\bth$ motivated in parts by the Gaussian prior assumption on $\bth$ and the fact that elliptical samplers do not require any tuning of the sampling scheme. Specifically, we use the adaptive generalized elliptical slice sampler \citep{marco2026adaptive} where a multivariate-t distribution is used for the auxiliary variable generation in defining the ellipses.

\subsection{Adaptation over localization threshold}
\label{sec:adapt}
The question of selecting an appropriate truncation level can be framed as a model selection question for which it is standard to use an information criterion, such as the well known Akaike information criterion (AIC) or Bayesian information criterion (BIC), to score each competing model and select the model with the best score. Most information criteria score a model by combining a numeric measure of the model's fit to the data with a numeric measure of model complexity. For example, AIC $=-2\log L(\hat\bth) + 2\cdot\dim(\bth)$ and BIC $= - 2\log L(\hat\bth) + \log n\cdot\dim(\bth) $, where $\hat\bth$ is typically the MLE of $\bth$. Models with low scores are considered better as they are able to explain the data better with lower model complexity. However, truncation level selection presents additional complications. 

First, notice that $\dim(\bth)$ does not change between two different choices of the truncation level. Consequently, it could be misleading to simply measure model complexity primarily by $\dim(\bth)$ as done in AIC and BIC. Given the Bayesian nature of our model fit, it could be more meaningful to use an alternative formulation such as the deviance information criterion (DIC) or the Watanabe-Akaike information criterion (WAIC) which uses an indirect measure of model complexity calculated from posterior uncertainty associated with the model fit. A precise formulation will be given shortly. Second, a deeper concern is that when the truncation level changes, the sample units that are used to fit the model can also change. Conceptually speaking, it is difficult or impossible to compare likelihood values obtained from two different data sets. Likelihood values are defined up to an arbitrary multiplicative constant which depends on the attributes of a particular data set in concern. When likelihood ratios are calculated based on the same data set, the arbitrary normalizing constant cancels, and the numeric value of the ratio remains interpretable. When ratios are calculated between two different data sets, no such cancellation takes place and the numeric value of the likelihood ratio becomes ambiguous. 

To mitigate the above two challenges we consider the following modification of WAIC:
\begin{equation}
\mathrm{WAIC}(M) = \underbrace{- 2\sum_{i \in S} \log \E_{\rm post}\{f(y_i|\bx_i;\bth,M)\}}_{\text{model fitness score}} + \underbrace{2\sum_{i \in S} \var_{\rm post}\{\log f(y_i|\bx_i;\bth,M)\}}_{\text{model complexity score}}
\label{eq:waic}
\end{equation}
where $M$ denotes a candidate model, $S$ is a subset of sample units that are common to all candidate model fits, and $\E_{\rm post}$ and $\var_{\rm post}$ refer to expectation and variance calculation with respect to the posterior distribution obtained by fitting model $M$. In our application, $S$ is simply the subset of sample units which are used to fit the model with maximum truncation ($\Delta = \frac1{10}$). The definition in \eqref{eq:waic} can be justified as follows. 
The usual WAIC score has close connection to leave-one-out cross-validation of Bayesian models \citep{watanabe2010asymptotic}. In the latter, posteriors are fitted in parallel to $n$ sub-samples obtained by removing each observation at a time and the log posterior predictive density is calculated for the left-out unit. For large samples, the sum total of these posterior predictive scores, times $-2$, is well approximated by the usual WAIC score which would replace the sum in \eqref{eq:waic} with a sum over all sample units. Likewise, the modified score in \eqref{eq:waic} considers leave-one-out posterior predictive scores only for units in the common subset $S$, providing a uniform test set for all candidate models in the analysis, while allowing each model to use different training sets to learn the posterior distribution of $\bth$. 

Intuitively,
if the global model was indeed correctly specified, one would expect all truncation levels to produce relatively similar numeric values for the first term in $\waic(M)$, while stronger localization levels (smaller $\Delta$), due to reduction in training sample size, will produce larger values for the second term, and hence the model with the best WAIC score is likely to be the one without any truncation. On the other hand, if the global model was misspecified, it would likely produce a larger numeric value for the first term in $\waic(M)$, indicating a mismatch between model assumptions and the reality around the threshold. Consequently, a model with a smaller $\Delta$ value may produce the best, i.e., the lowest WAIC score. 

The use of $\waic$ helps construct an adaptive version of our localized estimation method. First, fix a finite subset $D \subset (0,\min(t,1-t)]$ with a smallest element $\Delta^*$. Let $S^*$ be the subset of sample units $i$ with $|y_i - t| \le \Delta^*$. Then $S^*$ defines the largest subset of samples common to each truncation level $\Delta \in D$. Next, we fit the truncated version of \eqref{eq:par} restricted to the range $[t-\Delta,t+\Delta]$ for each $\Delta \in D$, and choose $\Delta = \hat \Delta$ with smallest WAIC score as calculated by \eqref{eq:waic} with $S = S^*$, and report the corresponding parameter estimates. In subsequent sections we use $D = \{\frac12, \frac25, \frac14,\frac1{10}\}$ for the case of $y \in (0,1)$ with $t = 1/2$. Note that $\Delta = \frac12$ corresponds to no localization. 


\section{Numerical experiments}
\label{sec:betareg-num}
To evaluate finite sample statistical performance of the proposed density discontinuity regression we carried out detailed numerical experiments with synthetic data generated from known distributions. The four major questions we seek to answer are:
\begin{enumerate}
\item When the parametric model matches the data-generating process, does the Bayesian method---applied to the complete data---provide accurate estimation and uncertainty quantification for $\balp$? 
\item How does the estimation quality degrade when the model is misspecified, i.e., when the baseline density $b(y | \bx)$ has a shape the beta family cannot represent? 
\item Does localization mitigate the misspecification bias and lead to improved estimation and uncertainty quantification?
\item Does the WAIC-based choice of $\Delta$ pick the right level of truncation and offer adaptive estimation accuracy as well as adequate statistical inference on $\balp$?
\end{enumerate}
Our results below show that the global parametric model performs well under correct specification (low bias, accurate coverage) but accumulates substantial bias when the model is misspecified relative to true outcome distributions that  are realistic in the context of threshold induced strategic manipulation. This finding reinforces the implications of the heuristic mathematical calculations in Section \ref{sec:theory} and sets up the need for data-adaptive localization. We also show below that the adaptive thresholding method from Section \ref{sec:adapt} answers the last two questions in the affirmative. 

\subsection{Experimental design}
We report results from a numerical experiment with 6 design conditions. For each condition, we simulated 200 data sets with $p = 6$ covariates, of which the first covariate $X_1$ was assigned a constant value of 1 for the intercept and the other covariates $X_2,\ldots,X_6$, were generated as independent standard normal draws. The outcome variable was taken to be bounded between zero and one, and the threshold was set to be $t = 1/2$. Half of these datasets had $n = 5000$ samples and the other half had $n= 20{,}000$ samples each. The six design conditions varied in the choice of the true jump coefficient vector $\balp_0$ (two choices) and the true baseline density $b_0$ (three choices). 

The two choices of the true jump coefficients were  
\begin{align*}
(\text{Large jump}) \qquad \balp_0 & =( 1.0, 0.3, 0.2, 0.2, 0.1,-0.1)',\\
(\text{Small jump}) \qquad \balp_0 & = ( 0.5, 0.2, -0.2, 0, 0, 0)'.
\end{align*} 
For the first choice, on average across replications, 99\% samples had a positive jump, with 50\% cases having $j(\bx) > 1$, i.e., $f(\tplus|\bx) > 2.7 f(\tminus|\bx)$. The jump size was most sensitive to changes in $X_2$, followed by $X_3$ and $X_4$, while $X_5$ and $X_6$ have mild impacts. For the second choice, jump prevalence was still high (96\%) but the median value of $j(\bx)$ was 0.5, and nearly 75\% cases had $f(\tplus|\bx) < 2 f(\tminus|\bx)$. However, only $X_2$ and $X_3$ had any impact on the jump magnitude, providing some clarity between relevant and irrelevant predictors. Accordingly, we refer to these two choices as ``Large jump'' and ``Small jump'', respectively.

The three choices of the true baseline density were  
\begin{align*}
(\text{Beta}) \qquad b_0(y|\bx) & = b^\star(y|\bx),\\
(\text{Near--beta}) \qquad b_0(y|\bx) & \propto b^\star(y|\bx) \cdot \exp\{\,\uI(y < t) \cdot (1 - e^{-\frac{(y-t)^2}{2\cdot (0.16)^2}})(\bx'\balp_0)_+\},\\
(\text{Beta-mix}) \qquad b_0(y|\bx) & = 0.5 \cdot b^\star(y|\bx) + 0.5 \cdot \frac{y^{14}(1 - y)^{9}}{B(15,10)},
\end{align*} 
where $b^\star(y|\bx)$ is the density of $\Be(s(\bx'\bgam^\star_1),s(\bx'\bgam^\star_2))$ with 
\[
\bgam^\star_{1} = (-1.5,-0.4,-0.1, 0.0, 0.4,-0.1)',~~\bgam^\star_2 = (-3.0,-0.1, 0.2,-0.6, 0.0,-0.1)'.
\] 
The first choice is an exact match for the beta parametric model of our estimation method and hence one would expect excellent statistical performance without any data trimming. The second and the third choices do not exactly match the beta baseline model and should require some level of truncation for good estimation accuracy. 

The six design choices present different degrees of challenge to our estimation methodology. In light of the asymptotic analysis presented in Section \ref{sec:theory}, we see that the three choices of the true baseline density correspond to $r = \infty$ for Beta, $r = 1$ for Near-beta, and $r = 0$ for Beta-mix. Therefore, an optimal MSE $\asymp n^{-1}$ could be obtained for the Beta baseline cases, with either choice of the jump coefficient, with no truncation whatsoever. The Near-beta cases should require moderate truncation ($\Delta \asymp n^{-1/5}$) to produce an optimal MSE $\asymp n^{-4/5}$. The Beta-mix cases should require the most aggressive truncation ($\Delta \asymp n^{-1/3}$) to produce an optimal MSE $\asymp n^{-2/3}$ matching the MSE of the optimal BOR method. 

While the asymptotic analysis gives an ordering of the optimal truncation levels as $n$ goes to infinity, it does not give a clear picture of what might happen in finite sample applications. For the latter, it is more useful to directly compare the true baseline $b_0$ to the model based formulation across different truncation levels $\Delta$. Table \ref{tab:kl} shows this comparison where the mismatch between the true baseline and the modeled baseline is measured by $\int \mathrm{KL}(b_0^\Delta(\cdot|\bx), b^\Delta(\cdot|\bx;\bgam)) dQ(\bx)$, the minimum of the $\bx$-averaged Kullback-Leibler divergence between the $t \pm \Delta$ localized versions of $b_0(\cdot|\bx)$ and $b(\cdot | \bx;\bgam)$. The smaller the number, the easier it is for the localized model to represent the localized true baseline, and hence less bias is incurred in estimation of $\balp$. We omit the Beta cases for which these localized KL divergence numbers are identically zero, and split the Near-beta cases by the corresponding choices of $\balp_0$ because they impact the true baseline $b_0$. Table \ref{tab:kl} confirms that the Beta-mix cases present the largest mismatch between the true and the model baseline densities, but these number improve quickly and steadily with the reduction of $\Delta$. Both Near-beta cases are somewhat easier, but shows slow improvement with the reduction of $\Delta$, until one gets to drastically small values. These observations will help explain the most salient findings of our numerical experiments which we report next.


\begin{table}[!t]
\centering
\caption{Local KL divergence between $b_0(\cdot | \mathbf{x})$ and its best Beta approximation
$\Be(s(\bx'\bgam^\dagger_1), s(\bx'\bgam^\dagger_2))$,
averaged over $\mathbf{x}$, for each trimming window $\Delta$.
``Drop'' is the absolute decrease in KL from $\Delta=\tfrac{1}{2}$ to $\Delta=\tfrac{1}{10}$;
``Ratio'' is $\mathrm{KL}(\Delta=\tfrac{1}{2})\,/\,\mathrm{KL}(\Delta=\tfrac{1}{10})$.}
\label{tab:kl}
\begin{tabular}{crrrrrr}
\toprule
& \multicolumn{4}{c}{KL divergence ($\times 100$) by $\Delta$} & \multicolumn{2}{c}{Sensitivity} \\
\cmidrule(lr){2-5} \cmidrule(lr){6-7}
Case & $\Delta=\tfrac{1}{2}$ & $\Delta=\tfrac{2}{5}$ & $\Delta=\tfrac{1}{4}$ & $\Delta=\tfrac{1}{10}$ & Drop & Ratio \\
\midrule
Near-beta $\times$ Large-jump & 0.30 & 0.28 & 0.23 & 0.02 & 0.27 &  12$\times$\\
Near-beta $\times$ Small-jump & 0.10 & 0.10 & 0.08 & 0.01 & 0.09 &  13$\times$\\
Beta-mix         			  & 8.14 & 4.84 & 0.53 & 0.04 & 8.10 & 209$\times$\\
\bottomrule
\end{tabular}
\end{table}

\subsection{Findings}
\begin{table}[!t]
\centering
\caption{Total MSE $\times 100$ of estimating $\boldsymbol{\alpha}$ for the six designs
and two sample sizes. Bold indicates the smallest MSE among the four fixed $\Delta$
choices within each row.}
\label{tab:mse_alpha}
\begin{tabular}{llr rrrr r r}
\toprule
Baseline & Jump & $n$
  & $\Delta\!\!=\!\!\tfrac{1}{2}$ & $\Delta\!\!=\!\!\tfrac{2}{5}$
  & $\Delta\!\!=\!\!\tfrac{1}{4}$ & $\Delta\!\!=\!\!\tfrac{1}{10}$
  & Adaptive & $\Delta$-Choice\%\\
\midrule
Beta      & Large & 5000  & \textbf{5.9}  & 6.7  & 11.7 & 73.6  & 6.9\ (1.2$\times$)  & 58/28/14/0  \\
          &       & 20000 & \textbf{1.7}  & 1.8  & 2.9  &  8.3  & 2.0\ (1.2$\times$)  & 58/28/13/1  \\
\cmidrule(l){2-9}
          & Small & 5000  & \textbf{18.1} & 25.2 & 74.8 & 230.8 & 55.8\ (3.1$\times$) & 48/30/12/10 \\
          &       & 20000 & \textbf{1.5}  & 1.6  & 9.6  & 21.1  & 1.7\ (1.1$\times$)  & 60/25/15/0  \\
\midrule
Near-beta & Large & 5000  & \textbf{10.0} & 10.8 & 12.0 & 89.8  & 18.9\ (1.9$\times$) & 28/27/40/5  \\
          &       & 20000 &  6.1 &  6.2 & \textbf{4.1}  &  7.1  & 4.5\ (1.1$\times$)  & 0/5/90/5    \\
\cmidrule(l){2-9}
          & Small & 5000  & \textbf{40.8} & 45.3 & 69.3 & 228.2 & 55.8\ (1.4$\times$) & 37/23/31/9  \\
          &       & 20000 & \textbf{2.1}  &  2.5 &  2.7 & 18.2  & 2.6\ (1.2$\times$)  & 27/25/48/0  \\
\midrule
Beta-mix  & Large & 5000  & 177.5 & 77.3 & \textbf{9.3}  & 13.3  & 10.8\ (1.2$\times$) & 0/0/86/14   \\
          &       & 20000 & 172.7 & 72.1 &  2.8 & \textbf{2.7}  & 3.0\ (1.1$\times$)  & 0/0/48/52   \\
\cmidrule(l){2-9}
          & Small & 5000  & 123.0 & 51.7 & \textbf{8.3}  & 67.7  & 28.9\ (3.5$\times$) & 0/0/83/17   \\
          &       & 20000 & 123.9 & 50.4 &  3.0 & \textbf{2.6}  & 2.6\ (1.0$\times$)  & 0/0/28/72   \\
\bottomrule
\end{tabular}
\end{table}

Table \ref{tab:mse_alpha} shows average accuracy of the proposed method in estimating $\balp$ at different levels of localization with a fixed $\Delta \in D = \{\frac12,\frac25,\frac14,\frac1{10}\}$ and also with the adaptive choice of $\Delta$ from $D$. Accuracy is measured as the total MSE---the total squared error $\|\hat\balp - \balp_0\|^2$ averaged across 100 replicated data sets---for each of the 6 design conditions and for each choice of the sample size $n$. We first draw some conclusions about the effect of localization on estimation accuracy, before commenting on the performance of the adaptive choice of $\Delta$.

For the Beta baseline cases, where the true baseline matches the model and hence no bias correction is necessary, localization only increased MSE by inflating variance. The Small-jump $\balp_0$ was harder to estimate at the smaller sampler size ($n = 5000$) but the difference went away at the larger sample size ($n=20{,}000$). For the Near-beta baseline, localization was helpful only in the case of Large-jump and at the larger sample size. In all other Near-beta cases, localization inflated total MSE though the changes were modest unless $\Delta = \frac1{10}$ was used. These results are in line with what we saw earlier in Table \ref{tab:kl}, the Near-beta baseline mismatch is overall quite mild and modest localization does not improve the match substantively. One needs to use a very small $\Delta$ to drastically reduce bias that arises from this baseline mismatch, and hence needs a large sample size so that variance also remains in control. However, for the Small-jump case, even $n = 20{,}000$ is not large enough to produce the right bias-variance tradeoff at $\Delta = \frac1{10}$. For the Beta-mix baseline, which presents a much larger mismatch between the truth and the model (Table \ref{tab:kl}), localization had the most dramatic effect. At sample size $n = 5000$, the choice of $\Delta = \frac14$ offered a $19\times$ improvement in total MSE for the Large-jump case, and $12\times$ improvement for the Small-jump case. At $n = 20{,}000$ the improvements were even more impressive at $64\times$ and $48\times$ reduction in total MSE respectively, with optimal $\Delta = \frac1{10}$. The overall conclusion is that the effect of localization runs the whole spectrum of being quite harmful to being extremely beneficial, depending upon the agreement between the true baseline and the model, the size of the jump, and the sample size. Consequently, adaptive localization is essential for the success of our parametric estimation methodology. 

Table \ref{tab:mse_alpha} shows that the adaptive localization technique---based on WAIC comparison---is quite effective across all cases for the larger sample size of $n = 20{,}000$. Indeed, at this sample size, the adaptively localized method estimated $\balp_0$ nearly as accurately as the method with the optimal (oracle) choice of $\Delta$, with at most a 20\% overall increase in the total MSE. However, its performance was much more variable at the smaller sample size of $n = 5{,}000$, with total MSE sometimes being 3.5 times of that corresponding to the optimal choice of $\Delta$, especially in the Small-jump cases.  

Beyond estimation accuracy, we assessed whether the adaptive method offered meaningful inference on the jump coefficients $\alpha_{0,j}$ associated with each covariate. We looked at four measures of performance of the estimate $\hat\alpha_j$ and the posterior credible interval $[\underline\alpha_j,\overline\alpha_j]$. Estimation bias (bias) was approximated by taking average of $\hat\alpha_j - \alpha_{0,j}$ over the 100 replicate data sets. Overall estimation accuracy was measured by the root mean square error (rmse) calculated as the square root of the average of $(\hat\alpha_j - \alpha_{0,j})^2$ across 100 replicates. Coverage of the credible interval (cvrg) was approximated by the proportion of 100 replicates in which we correctly had $\underline\alpha_j \le \alpha_{0,j} \le \overline\alpha_j$. We also looked at the rate of recovery of the sign of $\alpha_{0,j}$ (sign) by recording the fraction of replicates in which $[\underline\alpha_j,\overline\alpha_j]$ was on the same side of zero as the true non-zero $\alpha_{0,j}$, i.e., if either $\alpha_{0,j} > 0$ and $\underline\alpha_j > 0$ or $\alpha_{0,j} < 0$ and $\overline \alpha_j < 0$. The results are summarized in Table \ref{tab:adaptive}.

\begin{table}[!tp]
\centering
\caption{Adaptive localization results: bias, rmse, coverage (cvrg) and sign recovery (sign)
of the 95\% posterior credible interval for $\alpha_j$, for $n = 5000$ and $n = 20000$.}
\label{tab:adaptive}
\begin{tabular}{llr r rrrr rrrr}
\toprule
& & & & \multicolumn{4}{c}{$n = 5000$} & \multicolumn{4}{c}{$n = 20000$} \\
\cmidrule(lr){5-8} \cmidrule(lr){9-12}
Baseline & Jump & $j$ & $\alpha_{0,j}$
  & bias & rmse & cvrg & sign
  & bias & rmse & cvrg & sign \\
\midrule
\multirow{12}{*}{Beta}
  & \multirow{6}{*}{Large}
    & 1 & $1.0$ & $-0.01$ & 0.15 & 95 & 100 & $-0.01$ & 0.09 & 95 & 100 \\
  & & 2 & $0.3$ & $-0.00$ & 0.11 & 94 &  85 &  $0.00$ & 0.05 & 98 & 100 \\
  & & 3 & $0.2$ &  $0.01$ & 0.09 & 97 &  60 &  $0.00$ & 0.05 & 96 &  99 \\
  & & 4 & $0.2$ & $-0.01$ & 0.08 & 99 &  49 & $-0.01$ & 0.05 & 96 &  96 \\
  & & 5 & $0.1$ &  $0.00$ & 0.09 & 94 &  17 & $-0.01$ & 0.06 & 92 &  49 \\
  & & 6 & $-0.1$&  $0.01$ & 0.09 & 96 &  16 &  $0.00$ & 0.05 & 93 &  59 \\
\cmidrule(l){2-12}
  & \multirow{6}{*}{Small}
    & 1 & $0.5$ & $-0.23$ & 0.56 & 83 &  54 & $-0.00$ & 0.07 & 94 & 100 \\
  & & 2 & $0.2$ &  $0.06$ & 0.24 & 92 &  55 &  $0.00$ & 0.04 & 97 &  99 \\
  & & 3 & $-0.2$& $-0.04$ & 0.22 & 94 &  61 & $-0.01$ & 0.05 & 96 & 100 \\
  & & 4 & $0.0$ &  $0.00$ & 0.19 & 95 &  -- & $-0.00$ & 0.05 & 94 &  -- \\
  & & 5 & $0.0$ &  $0.02$ & 0.22 & 96 &  -- &  $0.01$ & 0.05 & 91 &  -- \\
  & & 6 & $0.0$ & $-0.06$ & 0.23 & 97 &  -- &  $0.00$ & 0.05 & 89 &  -- \\
\midrule
\multirow{12}{*}{Near-beta}
  & \multirow{6}{*}{Large}
    & 1 & $1.0$ & $-0.03$ & 0.21 & 95 &  96 &  $0.00$ & 0.10 & 94 & 100 \\
  & & 2 & $0.3$ & $-0.11$ & 0.21 & 63 &  44 & $-0.10$ & 0.12 & 48 &  97 \\
  & & 3 & $0.2$ & $-0.05$ & 0.13 & 90 &  28 & $-0.05$ & 0.08 & 78 &  76 \\
  & & 4 & $0.2$ &  $0.01$ & 0.21 & 91 &  47 & $-0.04$ & 0.07 & 90 &  76 \\
  & & 5 & $0.1$ &  $0.04$ & 0.14 & 89 &  30 &  $0.02$ & 0.07 & 90 &  60 \\
  & & 6 & $-0.1$&  $0.01$ & 0.14 & 88 &  23 &  $0.01$ & 0.05 & 91 &  46 \\
\cmidrule(l){2-12}
  & \multirow{6}{*}{Small}
    & 1 & $0.5$ & $-0.19$ & 0.52 & 85 &  68 &  $0.02$ & 0.08 & 87 & 100 \\
  & & 2 & $0.2$ & $-0.11$ & 0.26 & 86 &  29 & $-0.06$ & 0.08 & 67 &  89 \\
  & & 3 & $-0.2$& $-0.10$ & 0.30 & 81 &  56 &  $0.02$ & 0.06 & 88 &  96 \\
  & & 4 & $0.0$ &  $0.02$ & 0.18 & 95 &  -- &  $0.01$ & 0.06 & 92 &  -- \\
  & & 5 & $0.0$ &  $0.09$ & 0.23 & 92 &  -- &  $0.04$ & 0.06 & 88 &  -- \\
  & & 6 & $0.0$ & $-0.04$ & 0.21 & 89 &  -- & $-0.01$ & 0.04 & 96 &  -- \\
\midrule
\multirow{12}{*}{Beta-mix}
  & \multirow{6}{*}{Large}
    & 1 & $1.0$ &  $0.04$ & 0.13 & 91 & 100 &  $0.01$ & 0.08 & 89 & 100 \\
  & & 2 & $0.3$ &  $0.07$ & 0.14 & 91 &  90 &  $0.03$ & 0.08 & 80 & 100 \\
  & & 3 & $0.2$ &  $0.03$ & 0.13 & 91 &  52 &  $0.01$ & 0.06 & 95 &  93 \\
  & & 4 & $0.2$ & $-0.01$ & 0.11 & 95 &  43 & $-0.02$ & 0.07 & 91 &  90 \\
  & & 5 & $0.1$ & $-0.05$ & 0.14 & 84 &  12 & $-0.03$ & 0.08 & 84 &  24 \\
  & & 6 & $-0.1$&  $0.01$ & 0.14 & 92 &  14 &  $0.00$ & 0.06 & 96 &  40 \\
\cmidrule(l){2-12}
  & \multirow{6}{*}{Small}
    & 1 & $0.5$ & $-0.08$ & 0.39 & 85 &  84 &  $0.01$ & 0.06 & 92 & 100 \\
  & & 2 & $0.2$ &  $0.07$ & 0.20 & 79 &  75 &  $0.03$ & 0.07 & 87 &  93 \\
  & & 3 & $-0.2$& $-0.03$ & 0.16 & 90 &  52 & $-0.00$ & 0.07 & 89 &  91 \\
  & & 4 & $0.0$ & $-0.02$ & 0.12 & 97 &  -- & $-0.00$ & 0.05 & 98 &  -- \\
  & & 5 & $0.0$ & $-0.05$ & 0.19 & 86 &  -- & $-0.03$ & 0.08 & 81 &  -- \\
  & & 6 & $0.0$ & $-0.00$ & 0.14 & 96 &  -- & $-0.01$ & 0.06 & 94 &  -- \\
\bottomrule
\end{tabular}
\end{table}

Looking at the coverage probabilities in Table \ref{tab:adaptive}, we see that the adaptive method gave reasonable quantification of uncertainty for the Beta baseline cases, even though it preferred a localized version of the model fitting in roughly 40-50\% of the cases. One noticeable exception is the under-coverage for the intercept term at $n = 5{,}000$ under Beta baseline and small jump. Here, the adaptive estimate had a substantial negative bias, i.e., it frequently underestimated the overall jump size. While under-coverage is undesirable, it is somewhat reassuring to note that the adaptive estimation---by applying data localization---erred on the side of being conservative on estimating the jump size. Importantly, sign recovery was reasonably good for all non-intercept predictors excepting those with small impact on the jump size ($X_5$ and $X_6$) in the Large-jump case. However, by comparing the results between $n=5000$ and $n = 20000$, we do see that large improvements took place on all metrics with an increased sample size. 

Similar conclusions hold for the other true baseline choices, but here we see that under-coverage by the posterior interval persisted for certain coefficients even at the larger sample size of $n = 20000$, especially for the Near-beta cases. However, a closer inspection shows that the corresponding estimates were mostly conservative. For example, in the Near-beta, Large-jump case, the true value of $\alpha_2$ was 0.3. At $n = 20{,}000$, $\hat\alpha_2$ had a negative bias but a high sign recovery rate, meaning that in most of those cases where the posterior credible interval failed to include the true value $0.3$, it was situation entirely within the interval $(0, 0.3)$, and thus offered the right sign but with a smaller magnitude of the true effect of $X_2$ on the jump size.

Additionally, the adaptive method outperformed the BOR analysis in almost all cases of this study. Appendix \ref{app:c} reports results for the latter for $n=20{,}000$. At $\Delta = \frac1{10}$, BOR appears to suffer from large estimation bias, resulting in poor estimation accuracy and low coverage of its 95\% confidence intervals. A substantive reduction of the bias requires using an extreme localization threshold of $\Delta = \frac1{80}$, for which coverage guarantees also recover, but the precision of sign recovery diminishes drastically. By ignoring information on the shape of the local density, the BOR method appears unable to simultaneously offer low bias and high precision, even in large samples and across various possible choices of the trimming window.


\section{Analysis of corporate voting}

\label{sec:voting}


Shareholder votes on management-sponsored proposals are a central mechanism of corporate governance. When the share of votes in favor of a proposal exceeds a fixed threshold (typically 50\%), the proposal is implemented; otherwise it fails. Because the threshold is sharp and management has both the motive and a range of tools to influence the outcome---soliciting votes, restructuring or withdrawing proposals at risk of failing, timing announcements---the distribution of vote shares around the threshold carries information about the prevalence of strategic manipulation. \citet{malenko2016role} use a regression-discontinuity design to show that a negative recommendation from Institutional Shareholder Services (ISS) significantly reduces shareholder support, and this effect is large enough that management has a strong incentive to manipulate outcomes when ISS opposes a proposal. \citet{aggarwal2019power} document that proposal support varies systematically with firm characteristics including past stock performance and firm size, suggesting that the latitude for manipulation is unlikely to be uniform across firms.

Evidence of such sorting in shareholder voting is by now well established. \citet{listokin2008management} first documented that management-sponsored proposals are disproportionately likely to end just above the majority threshold. \citet{bach2019close} document the mirror-image pattern for shareholder-sponsored proposals, where narrow failures are in excess, and note the implication that close votes cannot routinely be treated as quasi-random. \citet{babenko2026management}, studying binding management proposals over 2003–2019, estimate a log-density ratio of 1.27 at the passage threshold and identify specific channels through which close outcomes are shaped, including proposal withdrawal, meeting adjournment, and supplemental solicitation filings. That paper also reports how the discontinuity varies across subsamples defined by analyst coverage, institutional ownership, board independence, firm size, and ISS recommendation, and observes that several of these characteristics are correlated with one another---analyst coverage with firm size in particular---so that a single-covariate split cannot isolate the contribution of any one of them. Their response is a sequential double sort. Supplying a principled and general version of that adjustment step, along with coefficient-style summaries and quantified uncertainty for it, is exactly what the framework developed in this paper is designed to do.

The corporate voting data analyzed here came from the Institutional Shareholder Services (ISS) Voting Analytics database\footnote{Our sample overlaps with, but is not the same as, that of \citet{babenko2026management}, a paper co-authored by one of us. That paper studies binding management proposals over 2003–2019, applies each proposal's own vote requirement, and excludes advisory categories---director elections, say-on-pay, auditor ratification---as well as dual-class firms. We use management-sponsored proposals over 2003–2015 at the majority threshold $t = 0.5$. Because the advisory categories exhibit substantially weaker discontinuities, our unconditional jump estimate is correspondingly smaller than theirs. Our purpose here is methodological: we use the setting to illustrate the regression framework, not to revisit the substantive governance conclusions of that paper.}. We considered records of all management-sponsored proposals that were initiated by U.S. firms during the period 2003 to 2015, and augmented these with standard company-related variables from the Standard \& Poor's Compustat database and The Institutional Brokers' Estimate System database. The resulting set of records consisted of 52 variables and 30,566 proposals. {The five covariates summarized in Table~\ref{tab:real-variables} were selected on the basis of the substantive considerations described above.} 
For each proposal, the response variable was the fraction of votes cast in favor of the proposal with the passing threshold $t = 0.5$. After dropping observations with missing covariates or unreliable response measurement, we were left with 19,741 samples.

\begin{table}[!t]
    \caption{Explanation of all variables in the corporate proposal voting data.
    \label{tab:real-variables}}
    \centering
    \begin{tabularx}{\textwidth}{l|X}
    \toprule
    Variable Name & Explanation \\
    \hline
    Response & Support fraction (for passing threshold $t = 0.5$)\\
    \hline
    \parbox[t][][t]{1.2in}{ISS Against\\ Recommendation} & Institutional Shareholder Services (ISS) is a company providing a subscription service that provides opinions on how shareholders should vote on each proposal. This is a binary variable that is 1 when ISS recommends voting against the proposal. \\
    \hline
    Analyst Coverage & The logarithm of one plus the number of equity analysts that actively track and publish opinions on a company and its stock. \\
    \hline
    Past Stock Return & Return of each stock over the previous year \\
    \hline
    \emph{Q} ratio & The Q ratio, also known as Tobin's Q, equals the market value of a company divided by its assets' replacement cost. \\
    \hline
    Firm Size & The logarithm of the book value of assets of the company\\
    \bottomrule
    \end{tabularx}
\end{table}

As a starting point, Figure~\ref{fig:f1-marginal} shows McCrary-style density-discontinuity tests applied separately to subsamples defined by each covariate: the binary ISS recommendation, and equal-thirds partitions of the three continuous covariates ranked in order. Read on the ratio scale that our estimand targets, these panels are less informative than they first appear. The estimated ratio $R/L$ is essentially unchanged across the ISS split (2.36 versus 2.32), non-monotone across firm-size terciles (2.28, 4.45, 1.57), and only mildly declining in Q. Analyst coverage shows the clearest marginal pattern, with the lowest-coverage tercile exhibiting the sharpest jump (3.11, against 2.08 and 2.19). The absolute difference $R - L$ is far larger for ISS-opposed proposals (0.92 versus 0.18), but this reflects the much higher overall density of such proposals near the threshold rather than a larger proportional jump---an illustration of why the log-ratio, and not the density difference, is the quantity of interest. These marginal patterns therefore motivate the analysis but cannot answer the questions we posed: they do not isolate the partial effect of any covariate, they carry no quantified uncertainty in coefficient form, and on the scale that matters they are close to silent for three of the four covariates. The regression analysis below addresses all three limitations.

{\begin{figure}[!t]
\centering
\includegraphics[scale=0.5]{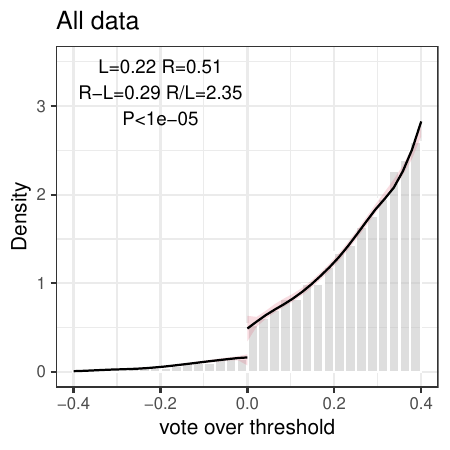}
\includegraphics[scale=0.5]{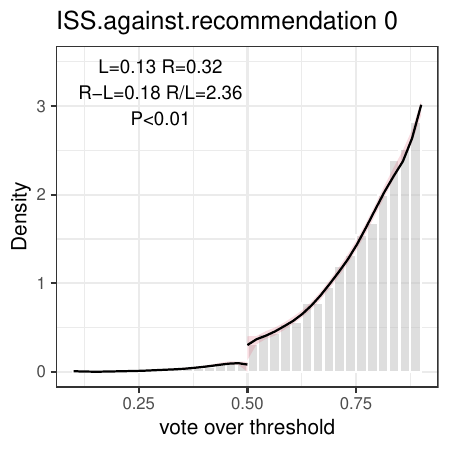}
\includegraphics[scale=0.5]{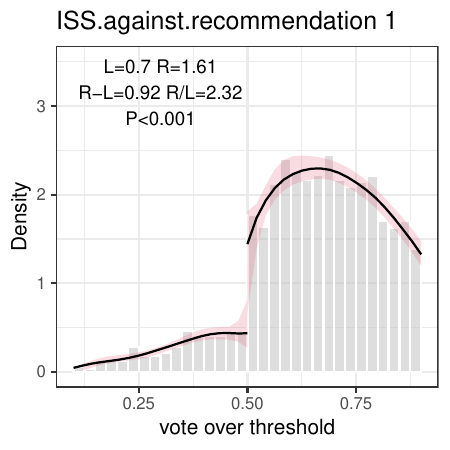}\\
\includegraphics[scale=0.5]{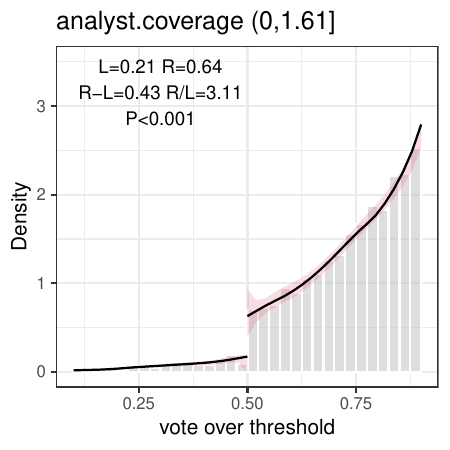}
\includegraphics[scale=0.5]{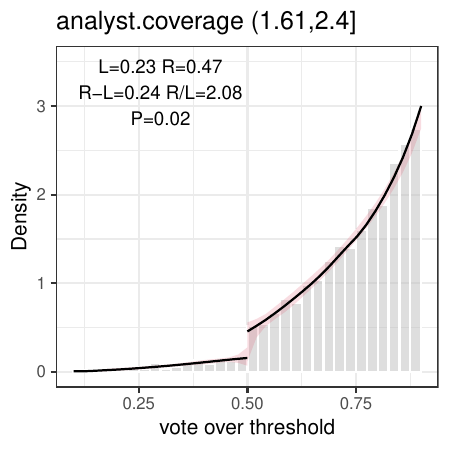}
\includegraphics[scale=0.5]{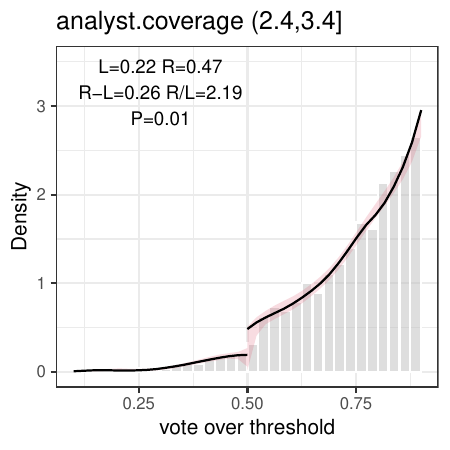}\\
\includegraphics[scale=0.5]{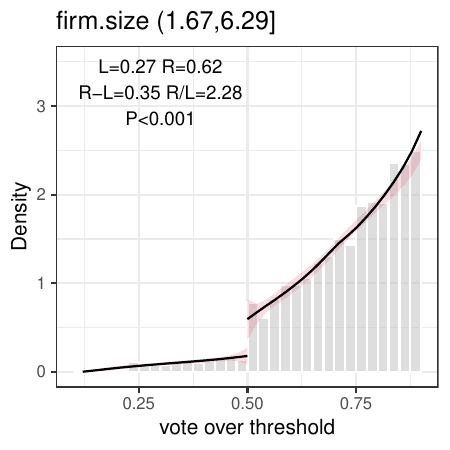}
\includegraphics[scale=0.5]{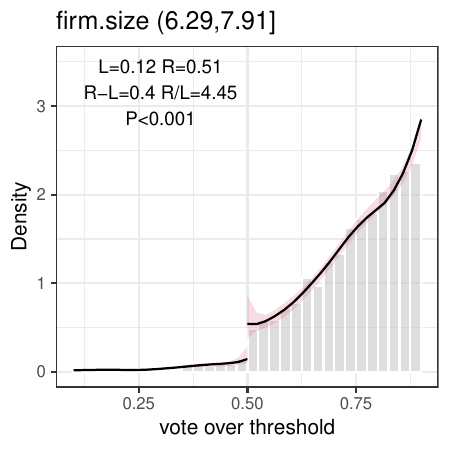}
\includegraphics[scale=0.5]{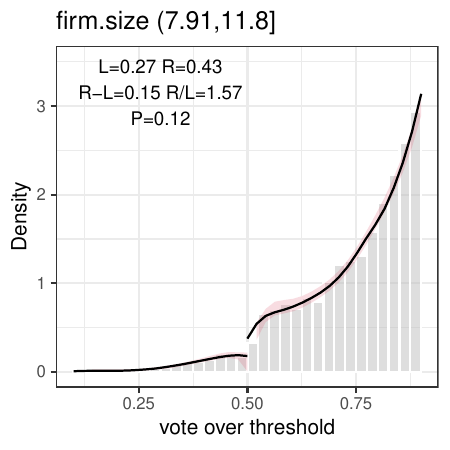}\\
\includegraphics[scale=0.5]{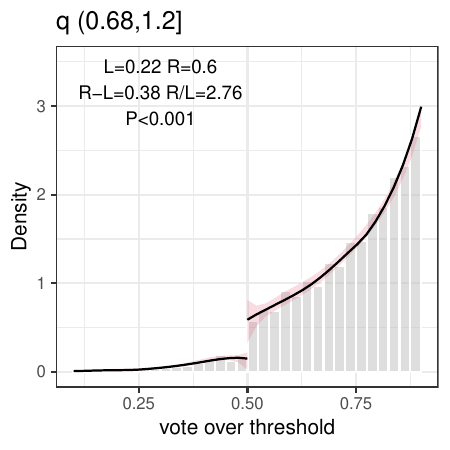}
\includegraphics[scale=0.5]{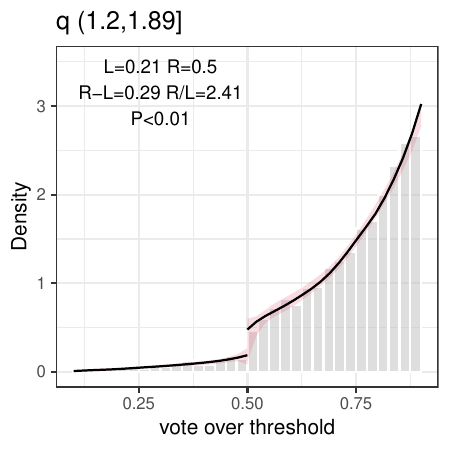}
\includegraphics[scale=0.5]{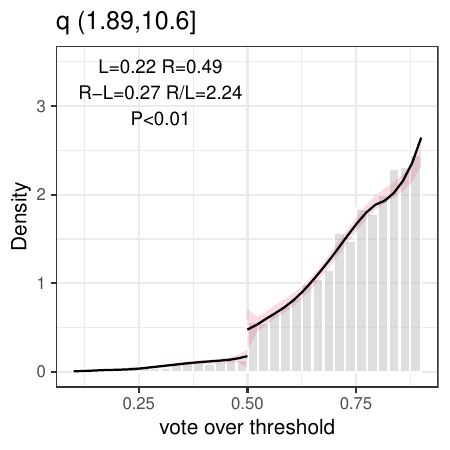}
\caption{Marginal density-discontinuity tests for the corporate voting data, segmented by covariate. ISS recommendation panels (top row) split the sample by the binary ISS-against indicator. Analyst coverage, firm size, and Q-ratio panels split the sample into equal thirds ranked by the covariate. Each panel reports the estimated left and right limits ($L$ and $R$), the difference $R-L$, the ratio $R/L$, and the discontinuity test $p$-value, computed via the \textsf{rddensity} R-package of \citet{cattaneo2020simple} with a cubic local polynomial fit.}
\label{fig:f1-marginal}
\end{figure}}

The data was analyzed with the adaptive estimation method described in Section \ref{sec:adapt}, with 4 levels of truncation $\Delta \in \{ \frac12,\frac25,\frac14,\frac1{10}\}$. Among these choices, moderate truncation ($\Delta = \frac14$) produced the smallest WAIC value (Table \ref{tab:vote}). All four WAIC values were calculated based on the largest subset of samples common to all truncation levels, namely the one corresponding to $\Delta = \frac1{10}$ with a sample size of 1,598. When broken down to its two parts in \eqref{eq:waic}, the model fit part kept improving with higher levels of truncation, but so did the measure of model complexity, with a substantial increase between $\Delta = \frac14$ and $\Delta = \frac1{10}$, offsetting a less dramatic improvement in the fitness measure between these two levels.

With $\Delta = \frac14$, the estimates of $\balp$ consisted of a positive intercept: $\hat\alpha_1 = 0.87$ with 95\% posterior credible interval = $(0.67,1.10)$, indicating a clear and persistent presence of jump discontinuity of the response density across the covariate space. A positive valued coefficient was estimated for the standardized indicator variable ``ISS against recommendation'' ($\hat \alpha_2 = 0.28$, 95\% interval = $(0.16,0.40)$).
In contrast, higher analyst coverage at the firm appeared negatively associated with jump discontinuity, with an estimated coefficient of $\hat \alpha_3 = -0.48$ (95\% interval = $(-0.68,-0.30)$). Neither past stock returns nor the Q ratio of the firm appeared strongly associated with density discontinuity. However, firm size, measured in logarithmic scale, appeared to have a positive association ($\hat \alpha_6 = 0.22$, 95\% interval = $(0.04,0.40)$).

Two features of these estimates serve as external checks on the method. First, because the non-intercept covariates are centered, $\hat\alpha_1 = 0.87$ is the model-implied log-density ratio at the mean covariate vector, corresponding to $R/L = 2.39$. The all-data panel of Figure \ref{fig:f1-marginal}, computed by an entirely different route---the boundary-adaptive local polynomial density estimator of \citet{cattaneo2024boundary}, with no parametric assumption on the baseline density and no covariates---gives $R/L = 2.35$, a log ratio of 0.85, comfortably inside our 95\% credible interval of (0.67, 1.10). The two quantities are not identical in population, since an unconditional density ratio averages densities rather than log-density ratios, but their close agreement indicates that neither the beta baseline nor the trimming window distorts the overall level of the estimated jump.

Second, the discrepancy between the binary-outcome estimates and ours is close to what the simulations predict. In the synthetic experiments of Appendix \ref{app:c} at $\Delta = \frac1{10}$, the logistic BOR intercept carries an upward bias of between 0.76 and 0.85 across all six design conditions. In the voting data at the same window, the BOR logistic intercept is 1.66 (Table \ref{tab:vote-simple}), against the nonparametric estimate of 0.85 from Figure \ref{fig:f1-marginal}---a discrepancy of 0.81---and against our own estimate of 0.91 (Table \ref{tab:vote}). The magnitudes need not coincide, since the BOR bias scales with the local slope of the log baseline density and this differs between the synthetic designs and the data. But the sign and approximate size of the discrepancy in the application are what the analysis of Section \ref{sec:theory} leads one to expect, and the agreement is closer than that analysis strictly guarantees.

As can be seen from Table \ref{tab:vote}, estimates of $\balp$ were somewhat similar across different levels of truncation with two notable exceptions. First, the effect of analyst coverage became stronger with higher levels of truncation. Second, the effect of firm size changed sign when some truncation was applied, though the effect size was not significant for any choice of $\Delta$ other than the one selected ($\Delta = \frac14$). The latter fact brings to light the critical need of choosing the truncation level in a meaningful and data-dependent fashion. 

 
\begin{table}[!t]
    \caption{Analysis of corporate voting data: estimates of $\alpha_j$, 95\% posterior credible interval (underneath in smaller fonts), and WAIC values are shown for various trim level $\Delta$ (broken underneath into the sum of model fitness and complexity scores). All covariates, including the ISS indicator, were standardized to have mean zero and variance 1.
    \label{tab:vote}}
\centering
\begin{tabular}{cc|rrrrrr|r}
\toprule
$\Delta$ & $n$ & \multicolumn{6}{c|}{Estimate of $\alpha_j$} & WAIC\\[3pt]
&& Intercept & ISS & Analyst & Return & Q-value & Size &\\
\hline
$\frac{1}{2}$ & 19,741  &  0.78    & 0.25  &   --0.30 & 0.08 & --0.05 & --0.12 & --5929\\[-6pt]
& & {\tiny (0.64,0.92)} & {\tiny (0.16, 0.36)} & {\tiny (--0.48, --0.15)} & {\tiny (--0.03,  0.20)} & {\tiny (--0.18,  0.08)} & {\tiny (--0.25,  0.03)} & {\tiny --5936+7}\\
$\frac{4}{10}$ & 11,455 & 0.84     & 0.22  &   --0.41 &  0.05 & 0.01 & 0.06 & --5937\\[-6pt]
& & {\tiny (0.66,1.00)} & {\tiny (0.10, 0.33)} & {\tiny (--0.57, --0.24)} & {\tiny (--0.08,  0.17)} & {\tiny (--0.13,  0.16)} & {\tiny (--0.12,  0.24)} & {\tiny --5947+10}\\
$\boldsymbol{\frac{1}{4}}$ & {\bf 5,155} &  \bf 0.87     & \bf 0.28  & \bf  --0.48 & \bf 0.02  & \bf 0.06 & \bf 0.22 & \bf --5946 \\[-6pt]
& & {\tiny \bf (0.67,1.10)} & {\tiny\bf  (0.16, 0.40)} & {\tiny\bf  (--0.68, --0.30)} & {\tiny\bf  (--0.11,  0.14)} & {\tiny\bf  (--0.10,  0.22)} & {\tiny\bf  (0.04,  0.40)} & {\tiny \bf --5958+12}\\
$\frac{1}{10}$ & 1,598 & 0.91     & 0.08  &   --0.56 &  0.19 & 0.09 & 0.13 & --5933 \\[-6pt]
& & {\tiny (0.50,1.20)} & {\tiny (--0.12, 0.27)} & {\tiny (--0.84, --0.26)} & {\tiny (--0.10,  0.45)} & {\tiny (--0.23,  0.37)} & {\tiny (--0.24,  0.44)} & {\tiny --5983+50}\\
\bottomrule
\end{tabular}
\end{table}

\begin{figure}[!t]
\centering
\includegraphics[scale=0.8]{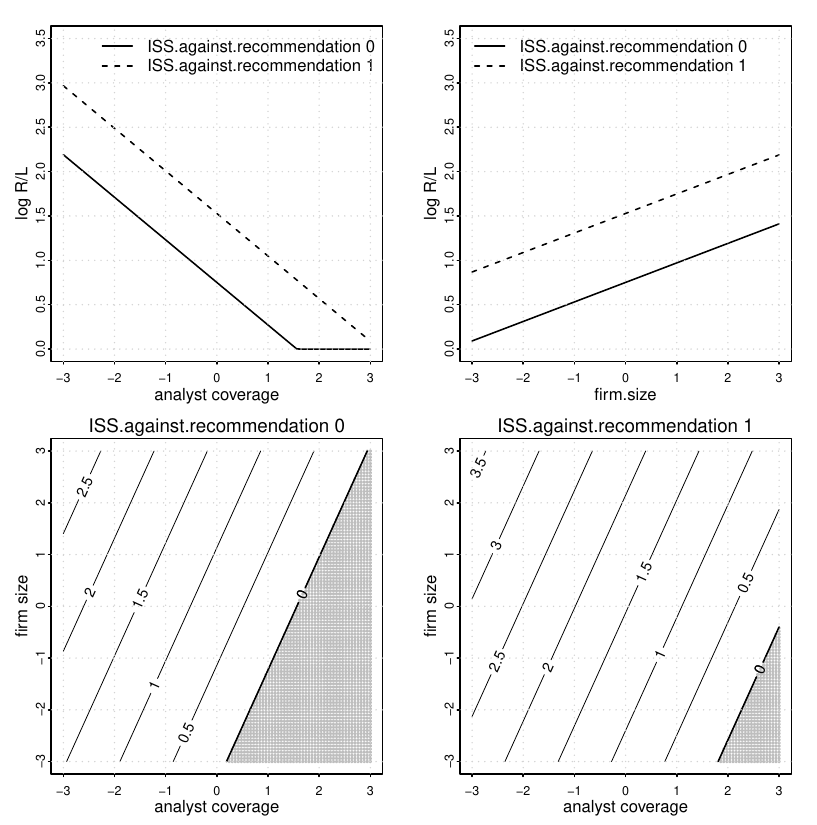}
\caption{Impact of ISS recommendation status, analyst coverage, and firm size on the size $j(\bx) = \log(R/L)$ of density discontinuity in corporate proposal voting. Analyst coverage and firm size are shown as z-scores, i.e.,  units of standard deviation departure from the mean. Top panels show estimated $\log (R/L)$ as function of ISS recommendation and either analyst coverage or firm size. Bottom panels show contours of $\log(R/L)$ as a joint function of analyst coverage and firm size separately for the two types of ISS recommendation. The shaded area is the region where $\log (R/L) = 0$, i.e., where no jump discontinuity manifests. 
}
\label{fig:vote-result}
\end{figure}

Several economic patterns emerge from these estimates, taking $\hat j(\bx) = (\bx'\hat\balp)_+$ with $\hat\balp$ from Table~\ref{tab:vote} for $\Delta = \frac14$; Figure~\ref{fig:vote-result} provides visual summaries, with $e^{\hat j(\bx)}$ directly comparable to the ratio $R/L$ from Figure~\ref{fig:f1-marginal}.

\emph{Manipulation is widespread but heterogeneous.} The estimated jump $\hat j(\bx_i)$ is positive for more than 99\% of proposals in the sample, indicating that strategic sorting near the majority threshold is essentially universal in management-sponsored proposals over our sample period. But its magnitude varies sharply with proposal and firm characteristics.

\emph{ISS opposition magnifies the jump.} On average, proposals receiving a favorable ISS recommendation have an estimated jump of $0.7$ (average of estimated $R/L = 2.2$); while the same figure for proposals opposed by ISS is $1.6$ (average of estimated $R/L = 5.6$). This is consistent with management responding most aggressively when ISS opposition signals a credible threat to passage, in line with the interpretation of \citet{malenko2016role}. The same ordering appears in the unconditional subsample comparison of \citet{babenko2026management}, whose binding-proposal sample yields log-density ratios of 1.40 for meetings with an ISS-against recommendation and 1.08 otherwise, though their indicator is defined at the meeting rather than the proposal level.

\emph{Analyst coverage attenuates the jump; firm size amplifies it.} A one standard deviation increase in analyst coverage is associated with a $0.5$ reduction in $\hat j(\bx)$, corresponding to a $40\%$ reduction in $R/L$ (conditional on the event that the jump is non-zero). A one standard deviation increase in firm size moves $\hat j(\bx)$ in the opposite direction by $0.22$, a $25\%$ increase in $R/L$. The contrasting signs are consistent with two distinct channels: external monitoring by analysts disciplines management, while larger firms have greater capacity for strategic sorting. The bottom panel of Figure~\ref{fig:vote-result} shows that combining both forces---ISS opposition, large firm size, low analyst coverage---yields estimated jumps as large as $j(\bx) = 3$ ($R/L = 20$). At the other extreme, manipulation is essentially absent for ISS-favored proposals from heavily covered firms, though such cases are rare in the data.

\emph{Adjustment matters for the firm-size finding.} Neither Tobin's $Q$ nor past stock return shows a detectable association with the discontinuity after adjustment. Firm size behaves quite differently. Marginally it shows no monotone pattern in our data (Figure \ref{fig:f1-marginal}), and in the binding-proposal sample of \citet{babenko2026management} the unconditional comparison runs in the opposite direction, with smaller firms exhibiting the larger discontinuity. Conditionally, however, the association is positive and clearly distinguishable from zero. The reversal has a transparent source. Firm size and analyst coverage are strongly positively correlated in our sample. An unconditional comparison across firm size therefore loads much of the analyst coverage effect onto size with the wrong sign. Holding analyst coverage fixed isolates the countervailing channel---larger firms have greater capacity for strategic sorting---that the marginal comparison masks. This is precisely the kind of finding a regression framework can deliver and a sample-splitting approach cannot, and it is the sense in which our estimates complement, rather than contradict, the subsample evidence in the prior literature.

\note{Additionally, our framework allows one to consider different modeling equations and compare them via WAIC to select a winning model. For example, Figure {\ref{fig:f1-marginal}}\note{} indicates that firm size may have a nonlinear effect on density discontinuity. To explore this possibility we fitted two variations of the above model, one which also included a quadratic effect of the firm size, and the other which replaced the linear term of firm size with non B-spline transformations of it (df = 3). On comparing WAIC values (for $\Delta = \frac14$, $n = 5,155$) we found that the B-spline formulation had the smallest WAIC score, followed by the original linear formulation, and the quadratic formulation was the worst. For the B-spline formulation, the estimated effect of analyst coverage was nearly indistinguishable ($\hat\alpha_3 = -0.47$, 95\% CI $=(-0.66,-0.25)$) from that in the original model, whereas the effect of ISS recommendation was slightly diminished ($\hat \alpha_2 = 0.22$, 95\% CI = $(0.08,0.36)$); see Figure \ref{fig:vote-result-firm-bspline} for more.}

\section{Concluding remarks}
\label{sec:discussion}

Density discontinuity tests have become a standard part of the applied econometrician's toolkit, but they answer only the first of two questions that researchers typically have: \emph{is there a jump?} and \emph{how does the jump vary with the institutional features that researchers can observe?} This paper formalizes the second question as a regression problem and develops a tractable estimator that addresses it directly. The framework targets the same log-density ratio that McCrary-style tests examine, now expressed as a function of covariates, with sign restrictions appropriate for one-sided manipulation and a data-driven choice of the local window around the threshold.

Three points are worth emphasizing in closing. First, the regression formulation makes adjustment matter in a way that segmentation analysis cannot. The firm-size finding in our corporate voting application---absent or reversed in unconditional comparisons, but statistically and economically meaningful once analyst coverage is held fixed---illustrates the value of holding other covariates fixed when interpreting heterogeneity in strategic behavior.
Second, the gain over binary outcome regression is real and traceable: by modeling the local density shape rather than ignoring it, we recover a faster convergence rate and avoid the sign-flip pathology that BOR produces in our application. Third, the sign restriction $j(\bx) \ge 0$ encodes economic content rather than just statistical convenience; it captures the directionality of strategic behavior in settings where the favorable side of the threshold is unambiguous, and it can be relaxed via $g(z) = z$ in two-sided settings such as union certification elections without any change to the computational machinery.

{The estimator we develop occupies a specific point on a tradeoff that runs through the paper. The regression formulation \eqref{eq:decomp} only requires a smooth baseline density $b(y \mid \bx)$, and Section~\ref{sec:theory} shows that, in principle, a sufficiently rich family for $b$ would let the estimator approach the parametric rate. Our choice of the beta family in \eqref{eq:par} caps that rate (Sections~\ref{sec:theory} and~\ref{sec:betareg}) in exchange for a closed-form normalizing constant, well-behaved posterior geometry, and tractable MCMC at moderate sample sizes. Section~\ref{sec:betareg-num} documents the resulting bias under a misspecified baseline density, our adaptive localization method mitigates this bias but may not eliminate it altogether. We view this combination as a defensible operating point given the inferential tools currently available; what remains is to ask what would be required to move along the frontier toward families with greater local fidelity.}

{The natural next step is a semiparametric extension of our model in which the baseline density $b(y|\bx)$ is modeled nonparametrically. Such an extension would in principle close the residual gap between the rate the beta family delivers and what richer families could deliver, since nonparametric density regression imposes essentially no restriction on the local degrees of freedom at $y = t$.} Two barriers stand in the way. The first is computational: introducing a jump discontinuity into a Bayesian nonparametric density regression requires evaluating a normalizing constant $Z(\bx)$ numerically for each of the $n$ observations at every likelihood call, which is roughly an order of magnitude more expensive than the beta formulation we adopt here. The second is methodological. Nonparametric Bayesian density estimation typically relies on infinite mixture representations, and these representations admit equivalent descriptions of the same density by very different cluster configurations. The resulting posterior multimodality complicates Monte Carlo inference: as evidence, three parallel MCMC runs of a logit stick-breaking density regression \citep{rigon2021tractable} on the corporate voting data explore different cluster-count regimes (Table~\ref{tab:lsbp}) while producing density estimates that agree broadly but differ in fine detail (Figure~\ref{fig:lsbp}). The methodological challenge is not the multimodality per se---which is well-understood \citep{dahl2005sequentially}---but ensuring that the discontinuity coefficient $\balp$ remains well-identified and adequately characterized by posterior summaries when the baseline density estimation problem is so loosely identified. {Resolving these barriers is, in our view, the principal direction along which the framework can be pushed; until it is, the parametric choice we adopt offers a workable operating point for the regression-style questions the paper is designed to answer.}

Beyond shareholder voting, the framework applies directly to other settings where strategic sorting around a known threshold is the substantive question of interest. Reported income distributions near tax-bracket cutoffs \citep{kleven2016bunching}, hedge-fund return distributions near zero \citep{bollen2009hedge}, and household income shares near the equal-earnings point \citep{bertrand2015gender} all share the same statistical structure as our application and admit the same regression-style questions about heterogeneity in the strength of strategic behavior. Implementing the method in any of these settings requires the researcher to make two choices---the local window $\Delta$ and the link function $g$---and the framework supplies a data-driven answer to the first and a clean institutional interpretation of the second.

 \section*{Acknowledgment} 
 We would like to thank Ilona Babenko, Abhiroop Mukherjee, George Panayotov, and Christopher D Walker for helpful comments on the manuscript. 


\bibliographystyle{chicago}
\bibliography{Bibliography}

\makeatletter
\renewcommand\thetable{A\@arabic\c@table}
\renewcommand \thefigure{A\@arabic\c@figure}
\makeatother

\appendix
\setcounter{figure}{0}  
\setcounter{table}{0}  

\section*{Appendices}


\section{Heuristic rate calculation}
\label{app:rate}

This appendix gives a formal statement and informal proof of the rate result of Section~\ref{sec:theory}. Throughout, $\dot s(u)$ denotes the derivative of a function $s(u)$, and we write $m_{0,l}(\bx) = \partial_y^l \log b_0(y \mid \bx)\vert_{y=t}$ for the $l$-th derivative at the threshold; similarly $m_l(\bx; \bgam) = \partial_y^l \log b(y \mid \bx; \bgam)\vert_{y=t}$ for the parametric family.

The localized maximum likelihood estimator is
\begin{equation*}
(\hat\bgam(\Delta), \hat\balp(\Delta)) \;=\; \arg\max_{\bgam, \balp} \prod_{i:|y_i - t| \le \Delta} f^\Delta(y_i \mid \bx_i; \bgam, \balp),
\end{equation*}
where $f^\Delta(y \mid \bx; \bgam, \balp)$ denotes the restriction of $f(y \mid \bx; \bgam, \balp)$ to $[t-\Delta, t+\Delta]$. Define the best Kullback--Leibler projection within the localized parametric model:
\begin{equation*}
(\bgam^\dagger(\Delta), \balp^\dagger(\Delta)) \;=\; \arg\min_{\bgam, \balp} \int d_{\mathrm{KL}}\bigl(f_0^\Delta(\cdot \mid \bx),\, f^\Delta(\cdot \mid \bx; \bgam, \balp)\bigr) q^\Delta(\bx)\, d\bx,
\end{equation*}
where $q(\bx)$ is the predictor distribution and $q^\Delta$ its restriction. 

Our calculations are carried out under three technical assumptions:
\begin{itemize}
\item[A1.] $b_0(t \mid \bx)$ is bounded away from zero for all $\bx$.
\item[A2.] There exists $\bgam_0 \in \Gamma$ such that $m_{0,l}(\bx) = m_l(\bx; \bgam_0)$ for $l = 1, \dots, r$.
\item[A3.] For all small $\delta > 0$, $\|\balp - \balp_0\|^2 \le K \int d_{\mathrm{KL}}\bigl(f_0^\delta(\cdot \mid \bx),\, f^\delta(\cdot \mid \bx; \bgam, \balp)\bigr) q(\bx)\, d\bx$ for all $\bgam, \balp$ and some universal constant $K$.
\end{itemize}

A1 is a mild boundedness condition standard in the density discontinuity literature. A2 is the substantive assumption: the first $r$ derivatives of $\log b_0(y \mid \bx)$ at $y = t$, viewed as functions of $\bx$, lie in the parametric family generated by $\bgam$. The integer $r$ measures the extent to which the parametric family captures the local behavior of $b_0$ at the threshold; $r = 0$ imposes no agreement and is vacuous. A3 is a curvature/identifiability condition relating Kullback--Leibler discrepancy to Euclidean discrepancy in $\balp$, standard in the literature on misspecified maximum likelihood and posterior contraction \citep{kleijn2006misspecification, van2008rates}.

\begin{proposition}[Rate of the localized MLE, heuristic]
\label{prop:rate}
Suppose A1--A3 hold, that the predictor support is bounded, and that the localized MLE satisfies the standard asymptotic normality of misspecified maximum likelihood with variance of order $(n\Delta)^{-1}$. Then for the localized MLE $\hat\balp(\Delta)$, with the optimal trimming choice $\Delta \propto n^{-1/(2r+3)}$,
\begin{equation*}
\E\{(\hat\alpha_j(\Delta) - \alpha_{0,j})^2\} \;\asymp\; n^{-(2r+2)/(2r+3)}, \qquad j = 1, \dots, p.
\end{equation*}
The rate is at least as fast as the BOR rate $n^{-2/3}$ (achieved at $r=0$) and {approaches the parametric rate $n^{-1}$ as $r \to \infty$. The largest $r$ for which A2 holds is a property of the parametric family rather than of the sample size, and for the beta family used in Section~\ref{sec:betareg} it is bounded by $r = 2$, giving an achievable ceiling rate of $n^{-6/7}$ (see Section~\ref{sec:betareg} for the calculation).}
\end{proposition}

\begin{proof}[Informal proof]
A1 implies that for every $y \in [t-\Delta, t+\Delta]$,
\begin{equation*}
f_0^\Delta(y \mid \bx) \;\asymp\; \frac{b_0(y \mid \bx)\, e^{-(\bx'\balp_0)\uI(y<t)}}{b_0(t \mid \bx)\{1 + e^{-\bx'\balp_0}\}\bigl\{1 + \sum_{l=1}^r \frac{\Delta^l}{l!} m_{0,l}(\bx) + O(\Delta^{r+1})\bigr\}},
\end{equation*}
and a similar expression holds for $f^\Delta(y \mid \bx; \bgam, \balp)$. Then, by A2,
\begin{equation*}
\sup_{|y-t| < \Delta} \left| \log \frac{f_0^\Delta(y \mid \bx)}{f^\Delta(y \mid \bx; \bgam_0, \balp_0)} \right| \;=\; O(\Delta^{r+1})
\end{equation*}
for every $\bx$. By Lemma 3.1 of \citet{van2008rates},
\begin{equation*}
\int d_{\mathrm{KL}}\bigl(f_0^\Delta(\cdot \mid \bx),\, f^\Delta(\cdot \mid \bx; \bgam_0, \balp_0)\bigr) q(\bx)\, d\bx \;=\; O(\Delta^{2(r+1)}).
\end{equation*}
Because $(\bgam^\dagger(\Delta), \balp^\dagger(\Delta))$ gives the best Kullback--Leibler projection, the same bound holds at the projection, and A3 gives $\|\balp^\dagger(\Delta) - \balp_0\| = O(\Delta^{r+1})$. Combining the squared bias $O(\Delta^{2(r+1)})$ with the assumed variance of order $(n\Delta)^{-1}$ gives
\begin{equation*}
\E\{(\hat\alpha_j(\Delta) - \alpha_{0,j})^2\} \;\asymp\; \Delta^{2(r+1)} + (n\Delta)^{-1},
\end{equation*}
and an optimal bias-variance tradeoff at $\Delta \propto n^{-1/(2r+3)}$ with rate $n^{-(2r+2)/(2r+3)}$.
\end{proof}

Three remarks are in order. First, the result is a rate calculation; the proportionality constants depend on $b_0$, the predictor distribution, and the parametric family, and we do not attempt to characterize them. Numerical assessment of practical bandwidth choice is taken up in Sections~\ref{sec:betareg-num}. Second, the variance assumption---that the localized misspecified MLE has variance of order $(n\Delta)^{-1}$---is the standard rate for likelihood estimators on a sample of effective size $n\Delta$, and we have not verified the conditions for it to hold uniformly in $\Delta$ as $\Delta \to 0$ and $n \to \infty$. A fully rigorous statement would require such uniform control; we leave this as a direction for future theoretical work. Third, A2 is an upper bound on what is attainable, not a claim about a fixed family. For a given parametric family, the largest $r$ for which A2 holds depends on $b_0$.

\newpage

\makeatletter
\renewcommand\thetable{S\@arabic\c@table}
\renewcommand \thefigure{S\@arabic\c@figure}
\makeatother

\appendix
\setcounter{figure}{0}  
\setcounter{table}{0}

\section*{Supplementary Material}

\section{Normalizing truncated beta distributions}
\label{app:a}
All modern computing platforms include efficient numerical routines to evaluate the ordinary beta function and the regularized incomplete beta function. However, in our statistical computation, we need to perform thousands of such evaluations simultaneously for each iteration of the Markov chain sampling. Such parallel evaluations can be too slow to carry out with typical implementations and it helps to use numerical routines that are geared toward vectorized evaluations. In the \texttt{R} programming environment vectorized evaluation of the beta function, in logarithmic scale, could be performed using the \texttt{Lbeta} function of the \texttt{Rfast} package. The regularized incomplete beta function $I_x(a,b)$, for $0 < x < 1$, could be written as the ratio $B(x,a,b)/B(a,b)$ where $B(x,a,b) = \int_0^x y^{a-1}(1 - y)^{b-1}dy$ is the incomplete beta function which can be further related to the hypergeometric function by the equation:
\[
B(x,a,b) = \frac{x^a (1 - x)^b}{a} \hypgeo{2}{1}(a+b,1,a+1,x).
\]
The \texttt{R} package \texttt{gls} offers vectorized evaluation of the hypergeometric functions. 

\section{Additional tables and figures}
\label{app:c}

Table \ref{tab:bor-20000} shows results of the binary outcome regression analysis (with $\Delta = \frac1{10}$ and $\Delta = \frac1{80}$) applied to synthetic data sets in our experiment ($n=20{,},000$). The truncated beta regression model we introduce in this work appears to universally dominates the the binary outcome regression method which fails to strike a reasonable tradeoff between bias and precision, either offering high bias and high precision (low coverage, high sign recovery) or low bias and low precision (high coverage, low sign recovery).

\begin{table}[!tp]
\centering
\caption{BOR analysis of synthetic data ($n = 20{,}000$).}
\label{tab:bor-20000}
\begin{tabular}{llr r rrrr rrrr}
\toprule
& & & & \multicolumn{4}{c}{$\Delta = \frac1{10}$} & \multicolumn{4}{c}{$\Delta = \frac1{80}$} \\
\cmidrule(lr){5-8} \cmidrule(lr){9-12}
Baseline & Jump & $j$ & $\alpha_{0,j}$
  & bias & rmse & cvrg & sign
  & bias & rmse & cvrg & sign \\
\midrule
\multirow{12}{*}{Beta}
  & \multirow{6}{*}{Large}
    & 1 &   1.0 &   0.82 & 0.82 &  0 & 100 &   0.13 & 0.29 & 89 &  99 \\
  & & 2 &   0.3 & --0.25 & 0.25 &  0 &  17 & --0.02 & 0.16 & 93 &  46 \\
  & & 3 &   0.2 & --0.17 & 0.17 &  8 &   6 & --0.02 & 0.15 & 94 &  26 \\
  & & 4 &   0.2 &   0.25 & 0.26 &  3 & 100 &   0.01 & 0.17 & 90 &  29 \\
  & & 5 &   0.1 &   0.28 & 0.28 &  1 & 100 &   0.02 & 0.16 & 95 &  11 \\
  & & 6 & --0.1 & --0.02 & 0.06 & 92 &  65 & --0.01 & 0.14 & 97 &  16 \\
\cmidrule(l){2-12}
  & \multirow{6}{*}{Small}
    & 1 &   0.5 &   0.82 & 0.82 &  0 & 100 &   0.12 & 0.23 & 91 &  92 \\
  & & 2 &   0.2 & --0.26 & 0.26 &  0 &   0 & --0.04 & 0.14 & 97 &  21 \\
  & & 3 & --0.2 & --0.15 & 0.16 & 13 & 100 & --0.04 & 0.14 & 93 &  44 \\
  & & 4 &   0.0 &   0.25 & 0.25 &  0 &  -- &   0.02 & 0.14 & 96 &  -- \\
  & & 5 &   0.0 &   0.30 & 0.30 &  0 &  -- &   0.04 & 0.13 & 98 &  -- \\
  & & 6 &   0.0 & --0.03 & 0.06 & 85 &  -- &   0.02 & 0.12 & 97 &  -- \\
\midrule
\multirow{12}{*}{Near-beta}
  & \multirow{6}{*}{Large}
    & 1 &   1.0 &   0.76 & 0.77 &  0 & 100 &   0.10 & 0.23 & 94 & 100 \\
  & & 2 &   0.3 & --0.27 & 0.28 &  0 &   6 & --0.02 & 0.15 & 95 &  45 \\
  & & 3 &   0.2 & --0.17 & 0.18 &  6 &   4 & --0.03 & 0.15 & 97 &  21 \\
  & & 4 &   0.2 &   0.24 & 0.25 &  1 & 100 &   0.02 & 0.17 & 92 &  31 \\
  & & 5 &   0.1 &   0.28 & 0.28 &  0 & 100 &   0.01 & 0.18 & 96 &   8 \\
  & & 6 & --0.1 & --0.02 & 0.05 & 94 &  67 & --0.00 & 0.15 & 95 &   8 \\
\cmidrule(l){2-12}
  & \multirow{6}{*}{Small}
    & 1 &   0.5 &   0.80 & 0.80 &  0 & 100 &   0.13 & 0.23 & 90 &  94 \\
  & & 2 &   0.2 & --0.27 & 0.27 &  0 &   1 & --0.03 & 0.15 & 92 &  22 \\
  & & 3 & --0.2 & --0.14 & 0.15 & 23 & 100 & --0.03 & 0.12 & 98 &  34 \\
  & & 4 &   0.0 &   0.25 & 0.26 &  0 &  -- &   0.03 & 0.14 & 93 &  -- \\
  & & 5 &   0.0 &   0.29 & 0.30 &  0 &  -- &   0.03 & 0.16 & 93 &  -- \\
  & & 6 &   0.0 & --0.04 & 0.06 & 90 &  -- & --0.00 & 0.14 & 88 &  -- \\
\midrule
\multirow{12}{*}{Beta-mix}
  & \multirow{6}{*}{Large}
    & 1 &   1.0 &   0.84 & 0.84 &  0 & 100 &   0.11 & 0.15 & 86 & 100 \\
  & & 2 &   0.3 & --0.07 & 0.08 & 59 & 100 & --0.00 & 0.10 & 96 &  88 \\
  & & 3 &   0.2 & --0.05 & 0.06 & 71 &  96 & --0.02 & 0.09 & 97 &  51 \\
  & & 4 &   0.2 &   0.08 & 0.09 & 43 & 100 & --0.01 & 0.10 & 92 &  56 \\
  & & 5 &   0.1 &   0.07 & 0.08 & 56 & 100 &   0.00 & 0.09 & 97 &  21 \\
  & & 6 & --0.1 & --0.01 & 0.04 & 96 &  79 & --0.00 & 0.10 & 96 &  20 \\
\cmidrule(l){2-12}
  & \multirow{6}{*}{Small}
    & 1 &   0.5 &   0.85 & 0.85 &  0 & 100 &   0.11 & 0.13 & 76 & 100 \\
  & & 2 &   0.2 & --0.07 & 0.08 & 45 &  98 & --0.02 & 0.08 & 96 &  59 \\
  & & 3 & --0.2 & --0.04 & 0.05 & 73 & 100 &   0.00 & 0.10 & 91 &  66 \\
  & & 4 &   0.0 &   0.08 & 0.09 & 24 &  -- &   0.01 & 0.08 & 92 &  -- \\
  & & 5 &   0.0 &   0.07 & 0.08 & 44 &  -- &   0.01 & 0.09 & 92 &  -- \\
  & & 6 &   0.0 & --0.01 & 0.03 & 95 &  -- &   0.01 & 0.08 & 93 &  -- \\
\bottomrule
\end{tabular}
\end{table}

\begin{figure}[!t]
\centering
\includegraphics[scale=0.9]{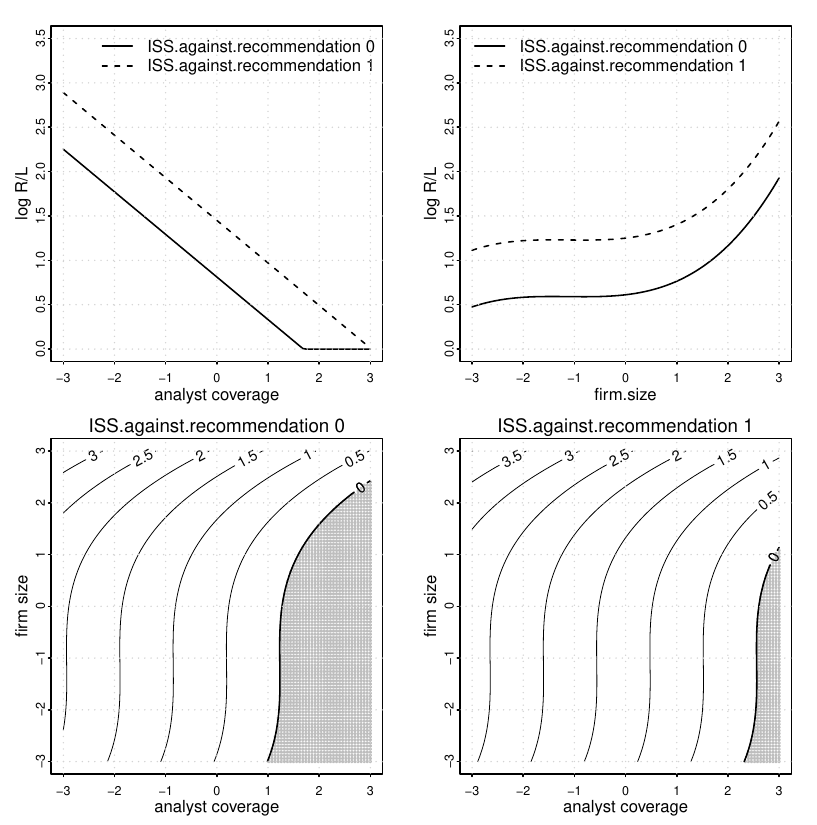}
\caption{A visual summary of the impact of ISS recommendation status, analyst coverage, and firm size on the size $j(\bx) = \log(R/L)$ of density discontinuity in corporate proposal voting where a nonlinear effect of firm size is accounted for through B-spline transformation.}
\label{fig:vote-result-firm-bspline}
\end{figure}

Figure \ref{fig:vote-result-firm-bspline} shows estimated effect of ISS recommendation, analyst coverage and firm size on density discontinuity in corporate voting outcomes, with a nonlinear effect of firm size modeled through B-spline transformations with 3 degrees of freedom. It appears that after adjusting for other variables, variation in firm size for small firms has little effect on density discontinuity, but very large firms appear to be associated with higher discontinuity.

\begin{figure}[!t]
    \centering
    \includegraphics[width=\linewidth]{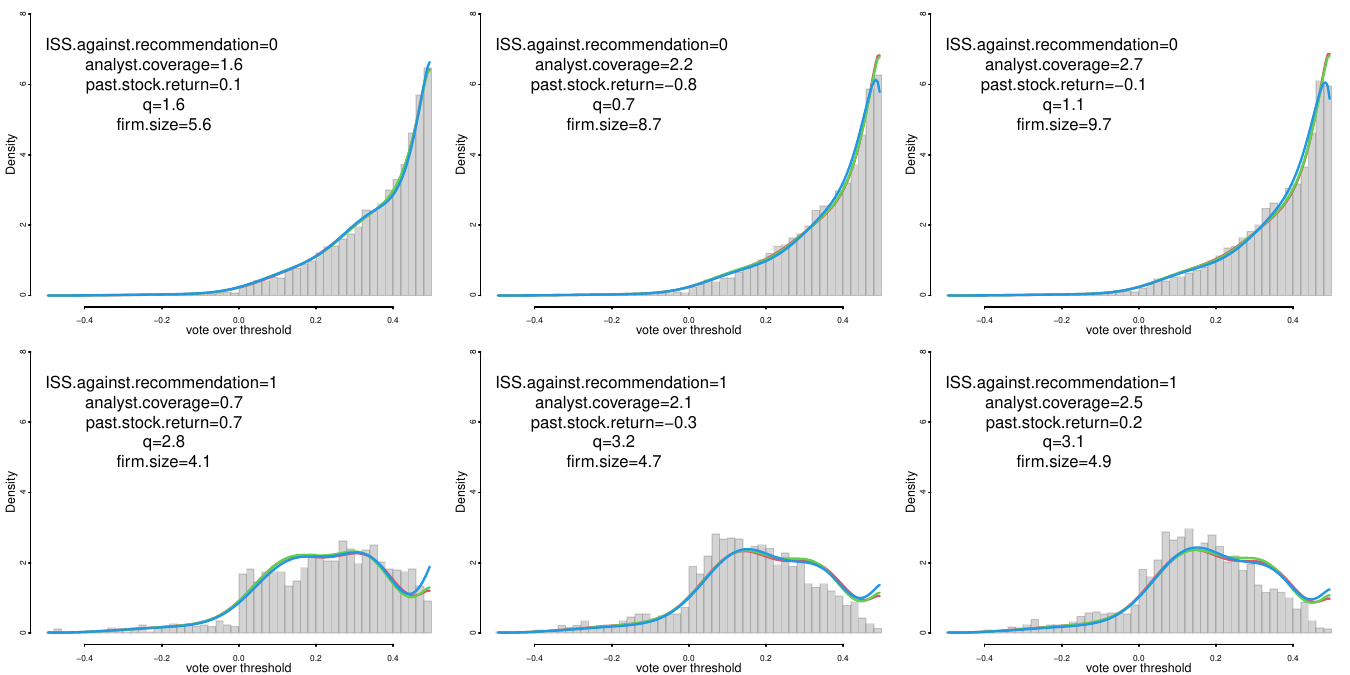}
    \caption{Estimates of $f(y|\bx)$ obtained from three separate runs of the Markov chain Monte Carlo algorithm for the logit stick-breaking process density regression model (with no jumps). Six panels correspond to six $\bx$ values associated with six different samples in the corporate voting data set. Panels on the top row have favorable ISS recommendation, whereas those on the bottom row have negative recommendation. Panels in the first column are firms with analyst coverage in the lower third of the distribution, those in the middle columns have moderate coverage, and those on the rightmost column have highest coverage. Other attributes, namely, past stock return, Q ratio, and firm size, are chosen randomly.}
    \label{fig:lsbp}
\end{figure}

Table \ref{tab:lsbp} demonstrates the problem of mixing of Markov chain samplers in exploring the posterior under the logit-stick breaking process model for (dependent) density estimation \cite{rigon2021tractable}. Shown are the posterior distribution of cluster counts obtained from three parallel runs of the same sampling algorithm, with different starting points and random number generation seeds. Maximum cluster count was restricted to be $H = 20$. The first and the third run explored clustering partitions with many components, where run 2 explored fewer number of clusters. Figure \ref{fig:lsbp} shows the estimated conditional density $f(y|\bx)$ for a few select $\bx$ values, superimposed with histogram of samples with comparable $\bx_i$ values. While broadly agreeing, the three estimates differ in finer details. Interestingly, runs 1 and 2, in spite of exploring different clustering configurations, produce similar density estimates. This observations underlines that the issue posterior multimodality for Bayesian mixture models is indeed a complex phenomenon. 

\begin{table}[!t]
    \caption{Cluster count distributions of 3 Markov chains for logit stick-breaking model. 
    \label{tab:lsbp}}
\centering
     \begin{tabular}{r|rrrrrrrrrrrr}
     \toprule
\#clusters &  6 & 7 & 8 & 9 & 10 & 11 & 12 & 13 & 14 & 15 & 16 & 17 \\
\midrule
Run 1 &  . & . & . & 1 & 3 & 14 & 33 & 27 & 10 & 7 & 4  & .\\
Run 2 &  1 & 3 & 9 & 20 & 35 & 20 & 9 & 1 & 1 & . & . & .\\
Run 3 &  . & . & . & .  & 5  & 16 & 26 & 27 & 19 & 2 & 3 & 1\\
\bottomrule
     \end{tabular}
\end{table}

\end{document}